\documentclass[aps,prd,reprint,superscriptaddress]{revtex4-2}
\usepackage{graphicx} 
\usepackage{amsmath}
\usepackage{derivative}

\usepackage[hidelinks]{hyperref}
\usepackage{orcidlink}

\usepackage{comment}

\usepackage{zref-clever}
\usepackage{booktabs}

\usepackage[acronym]{glossaries}
\glsdisablehyper
\makeglossaries
\newacronym{GW}{GW}{gravitational-wave}
\newacronym{NS}{NS}{neutron star}
\newacronym{BH}{BH}{black hole}
\newacronym{BBH}{BBH}{binary black hole}
\newacronym{BNS}{BNS}{binary neutron star}
\newacronym{BHNS}{BHNS}{black hole-neutron star}
\newacronym{EoS}{EoS}{equation of state}
\newacronym{EM}{EM}{electromagnetic}
\newacronym{sGRB}{sGRB}{short Gamma-Ray Burst}
\newacronym{ISCO}{ISCO}{Innermost Stable Circular Orbit}
\newacronym{CTT}{CTT}{Conformal Transverse-Traceless}
\newacronym{CTS}{CTS}{Conformal Thin Sandwich}
\newacronym{xCTS}{xCTS}{extended Conformal Thin-Sandwich}
\newacronym{ID}{ID}{initial data}
\newacronym{PDE}{PDE}{partial differential equation}
\newacronym{NR}{NR}{numerical relativity}
\newacronym{GR(M)HD}{GR(M)HD}{General Relativistic (Magneto) Hydrodynamics}
\newacronym{HRSC}{HRSC}{High-Resolution Shock-Capturing}
\newacronym{AMR}{AMR}{Adaptive Mesh Refinement}
\newacronym{GRHD}{GRHD}{general relativistic hydrodynamics}
\newacronym{GRMHD}{GRMHD}{general relativistic magnetohydrodynamics}

\usepackage{siunitx}
\DeclareSIUnit{\erg}{erg}
\DeclareSIUnit{\Mass}{\mathit{M}}
\DeclareSIUnit{\c}{\mathit{c}}
\DeclareSIUnit{\dyn}{dyn}
\DeclareSIUnit{\Gauss}{G}
\DeclareSIQualifier{\Sun}{\ensuremath{\odot}}
\DeclareSIQualifier{\disc}{disc}
\DeclareSIQualifier{\BH}{BH}
\DeclareSIQualifier{\NS}{NS}
\DeclareSIUnit\clight{\text{\ensuremath{c}}}
\DeclareSIUnit{\kms}{\kilo\metre\per\second}

\begin{document}

\title{RAKI: High-Order Alternative Finite Differences for General Relativistic Magnetohydrodynamics}

\author{Rahime Matur\,\orcidlink{0000-0001-9488-3817}}
\affiliation{University of Southampton, Southampton, United Kingdom}

\author{Beyhan Karakaş\,\orcidlink{0000-0002-3566-7329}}
\noaffiliation

\author{Ian Hawke\,\orcidlink{0000-0003-4805-0309}}
\affiliation{University of Southampton, Southampton, United Kingdom}

\author{Nils Andersson\,\orcidlink{0000-0001-8550-3843}}
\affiliation{University of Southampton, Southampton, United Kingdom}

\date{\today}

\begin{abstract}
    Binary neutron star (BNS) and black hole--neutron star mergers are among the most important multimessenger sources. Observations of these events could provide valuable insights into several unresolved phenomena, ranging from the efficiency of magnetic field amplification during and after mergers to the formation of magnetars and jets. Understanding these phenomena requires accurate modelling of magnetic field evolution, which remains a challenge for state-of-the-art general relativistic magnetohydrodynamics (GRMHD) simulations. In this work, we present \texttt{RAKI} (Relativistic Astrophysics Code Implementation of the Alternative Finite Difference Approach for GRMHD), the first GRMHD code to simulate BNS mergers using high-order alternative finite-difference methods with adaptive-order WENO (AFD-WENO-AO) reconstruction and a fully non-staggered, node-centred vector-potential formulation. The code achieves up to ninth-order convergence in smooth test problems while accurately capturing discontinuities. In magnetised Tolman--Oppenheimer--Volkoff (TOV) star tests, \texttt{RAKI} retains $99.2$ per cent of the initial maximum magnetic field strength over ${\sim}2$ ms, exhibiting an order of magnitude less magnetic field loss than a reference GRMHD code at the same resolution. In BNS merger simulations, fifth-order AFD-WENO-AO reconstruction yields approximately third-order convergence of the gravitational-wave phase in both magnetised and unmagnetised models. The discrete no-monopole constraint is preserved to machine precision without staggered grids, even during BNS mergers with moving refinement regions. These results establish \texttt{RAKI} as a high-order, non-staggered GRMHD framework for accurate modelling of magnetic field evolution in neutron star mergers and its implications for multimessenger astrophysics.
\end{abstract}

\maketitle

\section{Introduction}\label{sec:introduction}

Magnetic fields are generally expected to have little effect on the gravitational-wave signal during the inspiral phase of binary neutron star (BNS) mergers, but can significantly influence the remnant and its gravitational-wave and electromagnetic emission. The detection of GRB 170817A following GW170817 confirmed that BNS mergers can produce short gamma-ray bursts~\citep{2017PhRvL.119p1101A,doi:science_em170817}. As the neutron stars (NS) come into contact, strong shear at their interface triggers the Kelvin--Helmholtz instability (KHI), amplifying the magnetic field by orders of magnitude~\citep{price_rosswog_kh2006, anderson_2008_magbns, kiuchi_kh2015}. After merger, the magnetorotational instability (MRI) can further amplify the field~\citep{duez_magstar_2006, siegel_mri_2013, kiuchi_mri_2014}. In the remnant NS and accretion disc, magnetic stresses can redistribute angular momentum, affecting remnant stability and disc evolution~\citep{duez_magstar_2006, liu_magbns_remnant2008}. On secular timescales, magnetic fields can affect the properties of the ejecta, influencing the resulting kilonova emission~\citep{moesta_mag_kn2023}. Long-lived NS remnants with magnetar-strength magnetic fields may also launch relativistic jets~\citep{moesta_magnetar_2020}.

Numerical relativity simulations are required to study BNS mergers and their gravitational-wave (GW) and electromagnetic (EM) emission in a self-consistent manner. Most numerical relativity codes employ high-resolution shock-capturing methods based on finite-volume or finite-difference discretisations~\citep{whisky1_2003, whisky2_2005, whisky3_excision, Radice_thc::2012, Radice_thc2::2014, Radice_thc3::2014, Radice_thc4::2015, kiuchi_2012_grmhd, 2014CQGra..31a5005M, 2015CQGra..32q5009E, 2019MNRAS.490.3588M, 2020CQGra..37m5010C, 2025CQGra..42b5016K, 2025arXiv251106058F}. Alternative approaches include discontinuous Galerkin methods~\citep{2023CQGra..40b5004T, 2024CQGra..41x5002D} and smoothed particle hydrodynamics~\citep{2023arXiv230606226R}. These alternatives have specific advantages over standard grid codes, but introduce additional complexity compared to finite-volume and finite-difference approaches. Recent developments have also extended GRMHD simulations beyond the ideal-MHD approximation by incorporating finite electrical resistivity~\citep{2007MNRAS.382..995K, 2009MNRAS.394.1727P, 2025arXiv251018968A, 2025PhRvD.111l4033F, 2026arXiv260711670B, 2026arXiv260912998M}. 

The inclusion of magnetic fields introduces an additional numerical challenge, as the evolution must satisfy the no-monopole constraint, $\nabla\cdot\mathbf{B}=0$, imposed by Maxwell's equations. Different numerical approaches have been developed to handle this constraint. Some codes evolve the magnetic field directly and use \textit{divergence cleaning} to control violations of the no-monopole constraint~\citep{divclean_dedner_2002, Lehner_2010_divclean, 2014CQGra..31a5005M, 2024PhRvD.110h4046N, 2025CQGra..42x5005P}. Other codes employ \textit{constrained transport}~\citep{ct_evans_1988, toth_constrainttransport_2000, gardiner_2007}, which evolves the magnetic field on cell faces using electric fields calculated at cell edges~\citep{2025arXiv251106058F}. This approach preserves the discrete divergence of the magnetic field to machine precision. Another widely used method is to evolve the \textit{vector potential} and calculate the magnetic field from its curl~\citep{2015CQGra..32q5009E,2020CQGra..37m5010C,2025CQGra..42b5016K,2019MNRAS.490.3588M, 2025ApJS..277....3C}. Existing GRMHD implementations of this approach typically employ staggered grids, where different variables are stored at different spatial locations.

Several studies have aimed to improve the accuracy of magnetic field evolution~\citep{2007A&A...473...11D,2009MNRAS.394.1727P}. Among these, the ECHO framework employs high-order finite-difference methods with corrections to numerical fluxes, achieving up to fifth-order spatial accuracy in smooth test problems~\citep{2007A&A...473...11D}. The Frankfurt IllinoisGRMHD (FIL) code adopts a fourth-order implementation of the ECHO scheme~\citep{2019MNRAS.490.3588M}. 

In this work, we present \texttt{RAKI}, a new GRMHD code employing the alternative finite-difference (AFD) method with adaptive-order weighted essentially non-oscillatory (WENO-AO) reconstruction to achieve up to ninth-order spatial accuracy in smooth test problems. In \texttt{RAKI}, we evolve the vector potential on a non-staggered, node-centred grid, while preserving the no-monopole constraint to machine precision, even in binary neutron star simulations involving moving neutron stars and dynamical spacetimes.

Our CPU-based implementation supports both analytic and finite-temperature, composition-dependent equations of state, with conservative-to-primitive variable recovery performed using the \texttt{RePrimAnd} library~\citep{2021PhRvD.103b3018K,2022PhRvD.105j3031K}. The current version of \texttt{RAKI} solves the equations of ideal GRMHD; finite electrical resistivity and neutrino transport are left for future development.

We validate the code using one-dimensional Balsara shock tubes, circularly polarised Alfv\'en waves, two-dimensional magnetic loop advection, three-dimensional TOV~\citep{tolman_1939, oppenheimer_volkoff_1939} stars, and binary neutron star mergers. We demonstrate up to ninth-order convergence in smooth test problems and third-order convergence in the gravitational-wave phase of binary neutron star simulations. The vector-potential formulation also maintains the discrete divergence of the magnetic field at round-off levels in the tests presented here.

This paper is organised as follows. We first present the equations of relativistic magnetohydrodynamics in Sec.~\ref{sec:GRMHD}, followed by the numerical methods employed in this study in Sec.~\ref{sec:methods}, including a brief introduction to the alternative finite-difference approach and the numerical implementation. We then present the results of our one-, two-, and three-dimensional tests, including binary neutron star simulations, in Sec.~\ref{sec:tests}. Finally, we summarise our findings in Sec.~\ref{sec:conclusion}. Further details of the alternative finite-difference approach are provided in Appendix~\ref{sec:AFD_details}.

 Throughout this work, we adopt geometrised units in which $G=c=1$, unless otherwise stated.

\section{General Relativistic Magnetohydrodynamics}\label{sec:GRMHD}

The evolution of matter and magnetic fields in dynamical spacetimes is described by the equations of general relativistic magnetohydrodynamics (GRMHD). The governing equations consist of the conservation laws for rest mass and energy-momentum, together with the homogeneous Maxwell equations,
\begin{equation}
    \nabla_{\mu}(\rho u^{\mu}) = 0,\qquad
    \nabla_{\mu}T^{\mu\nu} = 0,\qquad
    \nabla_{\mu}{}^{*}F^{\mu\nu}=0,
\end{equation}
where $\rho$, $u^{\mu}$, and $T^{\mu\nu}$ denote the fluid rest-mass density, four-velocity, and energy-momentum tensor, respectively, and ${}^{*}F^{\mu\nu}$ is the dual Faraday tensor. In the absence of magnetic fields, the energy-momentum tensor of a perfect fluid is given by
\begin{equation}
    T^{\mu\nu} = \rho h u^{\mu}u^{\nu} + p_\mathrm{gas} g^{\mu\nu},
\end{equation}
where $h=1+\epsilon+p_\mathrm{gas}/\rho$ is the specific enthalpy, $\epsilon$ is the specific internal energy, $p_\mathrm{gas}$ is the fluid pressure, and $g^{\mu\nu}$ is the inverse spacetime metric.

When magnetic fields are included, the energy-momentum tensor becomes
\begin{equation}
    T^{\mu\nu}=(\rho h+b^2)u^{\mu}u^{\nu}+\left(p_\mathrm{gas} + p_\mathrm{ mag}\right)g^{\mu\nu}-b^{\mu}b^{\nu},
\end{equation}
where $b^\mu={}^*F^{\mu \nu}u_\nu$ is the magnetic field four-vector measured in the comoving fluid frame, and $b^2=b^\mu b_\mu$. The magnetic pressure is given by $p_{\mathrm{mag}}=b^2/2$.

Using the $3+1$ decomposition of spacetime, the covariant divergence of an arbitrary four-vector $V^{\mu}$ can be expressed as
\begin{equation}
    \nabla_{\mu}V^{\mu} = \frac{1}{\alpha\sqrt{\gamma}}\partial_{\mu}\left(\alpha\sqrt{\gamma}V^{\mu}\right),
\end{equation}
while energy-momentum conservation gives
\begin{equation}
    \partial_{\mu}\left(\alpha\sqrt{\gamma}T^{\mu}{}_{\nu}\right) = \alpha\sqrt{\gamma}\Gamma^{\lambda}_{\nu\mu}T^{\mu}{}_{\lambda}.
\end{equation}
Here, $\alpha$ is the lapse function, $\beta^i$ is the shift vector, $\gamma$ is the determinant of the spatial metric $\gamma_{ij}$, and $\Gamma^{\lambda}_{\mu\nu}$ denotes the Christoffel symbols. The fluid three-velocity $v^i$, measured by an Eulerian observer, and the Lorentz factor $W$ are defined as
\begin{equation}
    v^i = \frac{u^i}{W}+\frac{\beta^i}{\alpha},\qquad W=\alpha u^0=\frac{1}{\sqrt{1-\gamma_{ij}v^iv^j}}.
\end{equation}
We also define the transport velocity
\begin{equation}
    \tilde{v}^i=\alpha v^i-\beta^i.
\end{equation}
The magnetic field $B^i$ measured by an Eulerian observer is related to the comoving magnetic field four-vector through
\begin{equation}
    b^0=\frac{W}{\alpha}B^iv_i,\qquad b^i=\frac{B^i}{W}+b^0\tilde{v}^i.
\end{equation}

We adopt the Valencia formulation of general relativistic magnetohydrodynamics~\citep{Valencia::1991, Valencia::1997, Valencia::1999, 2000PhRvD..61d4011F}, with the conserved variables defined as
\begin{equation}
\begin{aligned}
    D&=\rho W,\\
    S_i&=(\rho h+b^2)W^2v_i-\alpha b^0b_i,\\
    \tau&=(\rho h+b^2)W^2-\left(p_{\mathrm{gas}} + p_{\mathrm{mag}}\right)-(\alpha b^0)^2-D.
\end{aligned}
\end{equation}

The balance law equations can then be written in flux-conservative form as
\begin{equation}\label{eq:hydroevolution}
\begin{aligned}
    &\partial_t(\sqrt{\gamma}D)+\partial_j(\sqrt{\gamma}D\tilde{v}^j)=0,\\
    &\partial_t(\sqrt{\gamma}S_i)+\partial_j\Bigl[\sqrt{\gamma}\bigl(S_i\tilde{v}^j+\alpha (p_\mathrm{gas} + p_\mathrm{mag})\delta^j_i - \alpha b_{i}B^{j}/W\bigr)\Bigr]\\  &\quad=\frac{\alpha\sqrt{\gamma}}{2}T^{\mu\nu}\partial_i g_{\mu\nu},\\
    &\partial_t(\sqrt{\gamma}\tau)+\partial_j\Bigl[\sqrt{\gamma}\bigl(\tau\tilde{v}^j+\alpha (p_\mathrm{gas} + p_\mathrm{mag}) v^j - \alpha^{2} b^{0} B^{j}/W\bigr)\Bigr]\\ &\quad=\alpha^2\sqrt{\gamma}\left(T^{0\mu}\partial_\mu\ln\alpha-\Gamma^0_{\mu\nu}T^{\mu\nu}\right),
\end{aligned}
\end{equation}
where $\delta^j_i$ is the Kronecker delta.

In ideal GRMHD, the electric field vanishes in the comoving fluid frame, and the dual Faraday tensor can be expressed as
\begin{equation}
    {}^{*}F^{\mu\nu}=u^{\mu}b^{\nu}-u^{\nu}b^{\mu}.
\end{equation}

The spatial components of the homogeneous Maxwell equations yield the induction equation
\begin{equation}
    \partial_t(\sqrt{\gamma}B^i)+\partial_j\left[\sqrt{\gamma}\left(\tilde{v}^jB^i-\tilde{v}^iB^j\right)\right]=0.
\end{equation}
The time component imposes the no-monopole constraint,
\begin{equation}
    D_iB^i=\frac{1}{\sqrt{\gamma}}\partial_i(\sqrt{\gamma}B^i)=0,
\end{equation}
where $D_i$ denotes the covariant derivative associated with the spatial metric. This constraint must be preserved throughout a numerical evolution.

As discussed in Sec.~\ref{sec:introduction}, different numerical methods have been developed to control violations of the no-monopole constraint, including divergence cleaning, constrained transport~\citep{ct_evans_1988, toth_constrainttransport_2000}, and the vector-potential approach~\citep{2003A&A...400..397D, zach_vectorpotential_2010}. In \texttt{RAKI}, we adopt the vector-potential approach and reconstruct the magnetic field from the curl of the spatial vector potential $A_i$:
\begin{equation}\label{eq:nomonopole}
    B^i=\frac{1}{\sqrt{\gamma}}\varepsilon^{ijk}\partial_j A_k,
\end{equation}
where $\varepsilon^{ijk}$ is the three-dimensional Levi-Civita symbol, with $\varepsilon^{123}=\varepsilon_{123}=+1$. Introducing the densitised magnetic field $\widetilde{B}^i=\sqrt{\gamma}B^i$, the no-monopole constraint becomes
\begin{equation}
    \partial_i\widetilde{B}^i=\varepsilon^{ijk}\partial_i\partial_jA_k=0.
\end{equation}
With compatible discrete curl and divergence operators, the no-monopole constraint can be maintained to round-off accuracy.

The electromagnetic four-potential is decomposed as
\begin{equation}
    \mathcal{A}_{\mu}=\Phi n_{\mu}+A_{\mu},
\end{equation}
where $n^{\mu}$ is the future-directed unit normal to the spatial hypersurfaces, $\Phi=-n^{\mu}\mathcal{A}_{\mu}$ is the scalar potential, and $n^{\mu}A_{\mu}=0$. Following~\citep{2015CQGra..32q5009E}, we adopt the generalised Lorenz gauge, which introduces a damping term to control gauge modes. The evolution equations for the vector potential and gauge scalar are
\begin{equation}\label{eq:magfield}
\begin{aligned}
    \partial_t A_i&=\varepsilon_{ijk}\tilde{v}^j\widetilde{B}^k-\partial_i(\alpha\Phi-\beta^jA_j),\\
    \partial_t(\sqrt{\gamma}\Phi)&=-\partial_j\left[\alpha\sqrt{\gamma}A^j-\beta^j(\sqrt{\gamma}\Phi)\right]-\xi\alpha\sqrt{\gamma}\Phi.
\end{aligned}
\end{equation}
We set $\xi=1.5/\Delta t_{\max}$ following \texttt{IllinoisGRMHD}~\citep{2003A&A...400..397D, 2005PhRvD..72b4028D, noble_2006igm, 2015CQGra..32q5009E}, where $\Delta t_{\max}$ is the timestep of the coarsest refinement level. 

In addition to the fluid and magnetic field evolution equations, we evolve the electron fraction $Y_e$ to track the composition of matter in neutron star merger simulations. Since weak interaction source terms are not currently included in \texttt{RAKI}, the electron fraction is passively advected with the fluid:
\begin{equation}\label{eq:efracevolution}
    \partial_t(\sqrt{\gamma}DY_e)+\partial_j(\sqrt{\gamma}DY_e\tilde{v}^j)=0.
\end{equation}

The fluid evolution equations can be written schematically in flux-conservative form as
\begin{equation}
    \partial_t U+\partial_jF^j=S,
\end{equation}
where $U$, $F^j$, and $S$ denote the evolved variables, fluxes, and source terms, respectively. For the fluid sector of our GRMHD system, the conserved state vector is given by~\citep{2000PhRvD..61d4011F}
\begin{equation}
    U_{\mathrm{fluid}}=\sqrt{\gamma}\left[D,\,S_i,\,\tau,\,DY_e\right].
\end{equation}
Together with the vector potential and generalised Lorenz gauge variables, the complete set of evolved variables is
\begin{equation}
    U=\left[\sqrt{\gamma}D,\,\sqrt{\gamma}S_i,\,\sqrt{\gamma}\tau,\,\sqrt{\gamma}DY_e,\,A_i,\,\sqrt{\gamma}\Phi\right].
\end{equation}

Finally, to close the system, an equation of state (EoS) is required to relate the thermodynamic variables. For an analytic EoS, the pressure can be expressed in terms of the rest-mass density and specific internal energy. For a finite-temperature, composition-dependent EoS, additional thermodynamic variables, including the temperature and electron fraction, are required. The numerical methods used to evaluate the fluxes, source terms, and time evolution are described in the following section.

\section{Numerical Methods}\label{sec:methods}

\subsection{Alternative Finite Difference (AFD)}\label{sec:AFD}

We consider high-order numerical methods for hyperbolic PDEs whose solutions may contain shock waves. In a series of papers~\cite{shuEfficientImplementationEssentially1988,shuEfficientImplementationEssentially1989}, Shu and Osher developed Essentially Non-Oscillatory (ENO) methods based on polynomial reconstruction of the relevant fields. These were subsequently generalised to Weighted ENO (WENO) methods~\citep{eno_harten1987, weno_shu_1994}.

When the system of PDEs forms a conservation or balance law such as
\begin{equation}
    \label{eq:pde1}
    \pdv{q}{t} + \pdv{f(q)}{x} = 0 \, ,
\end{equation}
consistency is typically enforced by using a flux-balance formulation, where the numerical fields are updated by ``inter-cell fluxes'', as
\begin{equation}
    \label{eq:update1}
    \odv{}{t} \hat{q}_i = \frac{1}{\Delta x} \left[ \hat{f}_{i-1/2} - \hat{f}_{i+1/2} \right] \, .
\end{equation}
Typically $\hat{q}_i$ is thought of as a cell average, and $\hat{f}_{i \pm 1/2}$ the inter-cell face-averaged fluxes. A standard approach in one spatial dimension is to reconstruct $q(x)$ as a piecewise-polynomial from the cell average values $\hat{q}_i$ to give two values at the cell faces, from which the inter-cell flux $\hat{f}_{i \pm 1/2}$ is computed from a (typically approximate) solution to the Riemann problem.

This way of thinking about the numerical solution is very useful, and provides direct links to the weak form of the PDE, helping us to understand the convergence behaviour of solutions involving discontinuities. However, it has significant practical problems when implemented beyond one spatial dimension. This follows from needing the face averaged inter-cell flux in the numerical integral. To compute this integral over the cell surface to high order, the number of reconstructions and Riemann problem solutions increases rapidly in higher dimensions.

The first alternative suggested by~\cite{shuEfficientImplementationEssentially1989} uses a finite difference reconstruction. It is important to note that the update formula Eq.~\ref{eq:update1} retains the flux-telescoping property (i.e., the cancellation of shared interface fluxes, ensuring discrete conservation and correct shock speeds), and therefore leads to consistent evolution of shocks, even when the terms are interpreted as node-centred. 
The idea is to reconstruct the flux $f(q)$ directly, so that the finite-difference approximation yields a high-order approximation to its derivative in any number of spatial dimensions. In the flux-splitting approach, the flux is separated into ``upwind'' and ``downwind'' components, which are reconstructed separately and then combined. This can reduce \emph{absolute} accuracy at low orders, but the method is straightforward to implement and extend to very high orders.

A second alternative  suggested by~\cite{shuEfficientImplementationEssentially1989} also takes a finite difference viewpoint, so still easily extends to high order in arbitrary spatial dimensions. However, it retains the idea of reconstructing the fields $q$ and solving the Riemann problem. In the finite difference context, however, the Riemann problem solution is not ``exact'' in any sense. Instead, it is the leading order approximation to the correct ``inter-cell flux'' that would give a high-order approximation to the derivative. Therefore, the idea is to write the intercell flux as, for example
\begin{equation}
    \label{eq:AFD_flux1}
    \hat{f}_{i+1/2} = f(q_{i+1/2}) - \frac{\Delta x^2}{24} \left. \pdv[2]{f}{x} \right|_{x=x_{i+1/2}} + \dots \, .
\end{equation}
The first term is precisely given by the solution of the Riemann problem at a single point. The higher order terms can be manipulated into Jacobian terms (as $J \sim \partial_q f$) and spatial derivatives of the fields $\partial_{x\dots x} q$, which can be directly computed from the reconstruction.

This approach was examined in detail by~\cite{merrimanUnderstandingShuOsher2003} and developed further by Balsara and collaborators~\cite[see, for example,][]{balsaraEfficientAlternativeFinite2024,balsaraEfficientFiniteDifference2024,balsaraAlternativeFiniteDifference2025a}, who refer to it as the Alternative Finite Difference (AFD or AFD-WENO) scheme. Compared with flux-splitting approaches, AFD-WENO can improve absolute accuracy while retaining a straightforward extension to high orders in multiple dimensions. Although the higher-order corrections make the formulation more involved, the studies of~\cite{balsaraEfficientAlternativeFinite2024,balsaraEfficientFiniteDifference2024} report computational costs comparable to those of flux-split WENO schemes.

The combination of pointwise Riemann solvers and high-order flux corrections provides a practical alternative to multidimensional face-averaged flux calculations. This forms the basis of the node-centred GRMHD implementation presented in this work. Full details of the implementation steps used here are given in~\zcref{sec:AFD_details}.

\subsection{Implementation}

We now describe the numerical implementation details used in our simulations, ranging from standard test problems to binary neutron star mergers.

As discussed in Sec.~\ref{sec:GRMHD}, we adopt the Valencia formulation~\citep{Valencia::1991, Valencia::1997, Valencia::1999, 2000PhRvD..61d4011F}, in which the evolution equations are written in flux-conservative form. The fluid evolution is therefore determined by evaluating the flux divergences and source terms.

We initialise each simulation with \gls{ID} appropriate to the problem under consideration. For the one- and two-dimensional tests, we developed the \gls{ID} thorn \texttt{Meze} by modifying the \texttt{GRHydro\_InitData}~\citep{whisky1_2003, whisky2_2005, whisky3_excision, 2014CQGra..31a5005M} and \texttt{GRHayLID}~\citep{grhayl} thorns provided by the \href{https://einsteintoolkit.org/}{\texttt{Einstein Toolkit}}~\citep{loffler_einstein_2012, zilhao_introduction_2013, EinsteinToolkit:web, etkhypatia} to meet the requirements of our numerical framework. We generate the initial data for the TOV tests using the \texttt{TOVSolver}. For the magnetised TOV tests and BNS simulations, we use the \texttt{Seed\_Magnetic\_Fields}, extended for the latter to measure the cylindrical radius relative to the centre of each star for the node-centred vector potential used in \texttt{RAKI}. We add a poloidal field confined to the stellar interior through the vector potential~\citep{2011PhRvD..83d4014G}
\begin{equation}
    A_\phi = \varpi^2 A_b \max(p - p_\mathrm{cut}, 0)^{n_\mathrm{s}},
    \label{eq:aphi}
\end{equation}
where $\varpi$ is the cylindrical radius measured from the centre of each star, $A_b$ sets the field strength, $p_\mathrm{cut} = 0.04\,p_\mathrm{max}$ confines the field to the region where $p > p_\mathrm{cut}$, and $n_\mathrm{s} = 2$. We choose $A_b$ such that the maximum field strength is $10^{15}\,\mathrm{G}$ at $t=0$, and compute $B^i$ from the curl of $A_i$.

Once the initial data have been specified, the evolution requires the recovery of primitive variables from the conserved quantities. The recovery of primitive variables from conserved quantities is commonly referred to as conservative-to-primitive (C2P) recovery. The conserved variables alone are not sufficient to evaluate the fluxes and source terms directly. For example, although the conserved rest-mass density $D$ is known, the corresponding $\rho$ must be recovered. The primitive variables are defined as
\begin{equation}
    \mathbf{p} = [\rho, v^{i}, p, \epsilon, B^{i}, Y_e].
\end{equation}
In general, this mapping (C2P) is not available in closed form and requires an iterative root-finding procedure, making the recovery numerically challenging and computationally expensive. The problem becomes more demanding when a temperature-dependent EoS is employed, as the thermodynamic variables must also be determined consistently during the recovery procedure; see~\citep{2018ApJ...859...71S} for further details.

In this work, we use the \texttt{RePrimAnd} library~\citep{2021PhRvD.103b3018K,2022PhRvD.105j3031K} to recover the primitive variables from the conserved ones. Our implementation is adapted from the minimal example distributed with the library~\footnote{\texttt{RePrimAnd}, \url{https://github.com/wokast/RePrimAnd} (accessed 29 July 2026)} and \texttt{AsterX}~\citep{2025CQGra..42b5016K}. Our implementation supports both analytic and finite-temperature, composition-dependent tabulated \gls{EoS}.

The next step is to calculate the numerical fluxes and their spatial derivatives. Although both exact and approximate Riemann solvers are available, approximate solvers are generally preferred in numerical relativity due to their lower computational cost. In classical finite-difference WENO schemes based on flux splitting, the fluxes are evaluated at grid points and split into positive- and negative-going contributions before reconstruction. This approach does not directly employ an interface Riemann solver. In contrast, finite-volume and alternative finite-difference methods reconstruct the primitive variables rather than the split fluxes. The resulting left and right states at each interface are then supplied to an approximate Riemann solver, allowing greater flexibility in its selection.

We employ the HLLE approximate Riemann solver~\citep{doi:10.1137/1025002} within the alternative finite-difference framework. More sophisticated Riemann solvers, such as HLLC~\citep{hllc_newt_2005, hllc_mignone_2006} or HLLD~\citep{hlld_newt_2005, hlld_mignone_2009}, could also be incorporated in future work without modifying the reconstruction procedure. The HLLE numerical flux at the interface $i+1/2$ is given by
\begin{widetext}
\begin{equation}\label{eq:fluxHLLE}
    \mathbf{F}^{\mathrm{HLLE}}_{i+1/2} = \frac{b^+_{i+1/2}\mathbf{f}(\mathbf{u}^L_{i+1/2})-b^-_{i+1/2}\mathbf{f}(\mathbf{u}^R_{i+1/2})+b^+_{i+1/2}b^-_{i+1/2}(\mathbf{u}^R_{i+1/2}-\mathbf{u}^L_{i+1/2})}{b^+_{i+1/2}-b^-_{i+1/2}},
\end{equation}
\end{widetext}
where $b^+_{i+1/2}$ and $b^-_{i+1/2}$ denote estimates of the fastest right- and left-propagating characteristic speeds, respectively, with $b^+_{i+1/2}\geq 0$ and $b^-_{i+1/2}\leq 0$. The quantities $\mathbf{u}^L_{i+1/2}$ and $\mathbf{u}^R_{i+1/2}$ are the reconstructed left and right states at the interface, and $\mathbf{f}(\mathbf{u})$ denotes the corresponding physical flux. In the AFD framework, the interface fluxes are subsequently supplemented by the high-order correction terms described in Sec.~\ref{sec:AFD}.

In addition to computing the numerical fluxes, the evolution of the magnetic field requires the reconstruction of $B^i$ from the vector potential. As described in Sec.~\ref{sec:GRMHD}, we evolve $A_i$ rather than $B^i$ directly. The densitised magnetic field is therefore reconstructed from the curl of the vector potential:
\begin{equation}
\begin{aligned}
    \widetilde{B}^x=&\partial_y A_z-\partial_z A_y,\qquad \\
    \widetilde{B}^y=&\partial_z A_x-\partial_x A_z,\qquad \\
    \widetilde{B}^z=&\partial_x A_y-\partial_y A_x.
\end{aligned}
\end{equation}
For nominal third-, fifth-, seventh-, and ninth-order runs, these pointwise derivatives are evaluated using fourth-, sixth-, eighth-, and tenth-order centred finite-difference operators, respectively. The curl operator therefore has a higher formal spatial order than the corresponding evolution scheme and does not introduce a lower-order truncation error in smooth regions. For example, when fifth-order WENO reconstruction is employed, the spatial derivative appearing in the curl operator is approximated using the sixth-order centred finite-difference formula
\begin{equation}
\begin{aligned}
    f'_i = \frac{1}{60\Delta x}\bigl(
    &-f_{i-3}+9f_{i-2}-45f_{i-1}\\
    &+45f_{i+1}-9f_{i+2}+f_{i+3} \bigr) +O(\Delta x^6).
\end{aligned}
\label{eq:derivative6}
\end{equation}
We next highlight several implementation choices adopted in \texttt{RAKI}.

\begin{enumerate}
    \item We calculate the Jacobian matrix in separate blocks rather than constructing and storing the entire matrix in a single large array. Specifically, we compute the $6\times6$ block corresponding to the fluid variables and the remaining $4\times4$ block separately. This allows us to evaluate the two blocks independently without constructing or storing the full $10\times10$ Jacobian matrix.

    \item We use component-by-component WENO-AO interpolation rather than characteristic WENO-AO interpolation. As noted by~\cite{balsaraEfficientAlternativeFinite2024}, the component-by-component approach is computationally less expensive while maintaining the same formal order of accuracy in smooth regions. We have also implemented the characteristic version and found it to be significantly more computationally expensive, consistent with the findings of~\cite{balsaraEfficientAlternativeFinite2024}.

    To improve robustness near discontinuities, we employ the Balsara--Kim flattener~\citep{2016JCoPh.312..357B} in the Balsara shock-tube, magnetic loop advection, and binary neutron star simulations. In addition, the Suresh--Huynh monotonicity-preserving limiter~\citep{1997JCoPh.136...83S} is applied at hydrodynamic cell interfaces. 

    \item In the periodic one-dimensional tests, the magnetic field is reconstructed at each time step from the evolved vector potential according to
    \begin{equation}
        \mathbf{B}=\mathbf{B}_0+\boldsymbol{\nabla}\times\mathbf{A},
    \end{equation}
    where $\mathbf{B}_0$ represents the constant background magnetic field. 

    \item We employ Kreiss--Oliger (KO) dissipation for the evolution of the vector potential to suppress high-frequency numerical oscillations.
\end{enumerate}

Having described the reconstruction and flux calculation procedures, we now turn to the geometrical source terms that couple the fluid evolution to the spacetime geometry provided by the \texttt{Einstein Toolkit}. We compute the components of the energy-momentum tensor and store them through the \texttt{TmunuBase} interface. The Valencia momentum source terms require the first spatial derivatives of the spacetime metric, whereas the energy source term contains contractions with the extrinsic curvature and the first spatial derivatives of the lapse. Consequently, the source-term calculation requires $\partial_i\alpha$, $\partial_i\beta^j$, and $\partial_i\gamma_{jk}$, together with the extrinsic curvature $K_{ij}$. All required spatial derivatives are evaluated using second-, fourth-, sixth- and eighth-order centred finite-difference stencils. The geometrical source terms are evaluated after the flux divergence and are subsequently added to the right-hand sides of the conservative evolution equations.

We use the \texttt{Carpet} adaptive mesh refinement (AMR) driver~\citep{2004CQGra..21.1465S, CarpetCode:web} within the Cactus framework~\citep{allen_cactus_1999, goodale_cactus_2003, Cactuscode:web, Cactusprize:web}. The coupled spacetime and magnetohydrodynamics equations are evolved using the Method of Lines (MoL).

\section{Tests}\label{sec:tests}

To validate our code, we perform a series of one-, two-, and three-dimensional tests to assess its ability to capture the expected behaviour in both smooth regions and across discontinuities, achieve the expected order of accuracy, and satisfy the no-monopole constraint to machine precision.

\subsection{1D Tests}

In this subsection, we assess the accuracy and robustness of the numerical methods using circularly polarised Alfv\'en waves and magnetised shock-tube tests. The former allows us to measure the convergence order of the schemes in smooth flows, while the latter examines their ability to capture shocks and discontinuities.

\subsubsection{Circularly Polarised Alfv\'en Waves}

High-order numerical methods can produce spurious oscillations near discontinuities. To control these oscillations, we employ WENO-AO reconstruction~\citep{balsaraEfficientAlternativeFinite2024}, which adapts the local approximation according to the smoothness of the solution. Since discontinuities can reduce the observed convergence order, we first assess the accuracy of our numerical schemes using circularly polarised Alfv\'en (CP-Alfv\'en) waves~\citep{1989cup..book.....A,1997PhLA..232..435K,2007A&A...473...11D}.

The computational domain is defined by $(x,y,z)\in[-0.5,0.5]^3$, with the wave propagating along the $x$-direction and a wavelength of unity. Periodic boundary conditions are imposed in all spatial directions. We set $\rho=1$, $p=0.5$, and $\Gamma=5/3$, with a longitudinal magnetic field $B^{x}=1$ and velocity $v^x=0$. We also set $B^{y}= B^{x}\cos(kx)$, where $k = 2\pi/L$ is the wavenumber and $L$ is the wavelength, corresponding to the length of the periodic domain ($L = 1$). The corresponding Alfv\'en speed is $v_A=0.5$. All simulations are evolved until $t_{\mathrm{final}}=2$, corresponding to one wave period. We employ the third-, fifth-, seventh-, and ninth-order AFD-WENO-AO schemes for both the fluid and magnetic field evolution, with resolutions $N_x=16,32,64,$ and $128$, and an additional $N_x=96$ for the ninth-order scheme, while keeping $N_y=N_z=8$. The time step is reduced with increasing resolution to suppress temporal truncation errors at all orders in order to measure the spatial convergence order.

\begin{figure}
    \centering
    \includegraphics[width=1.0\linewidth]{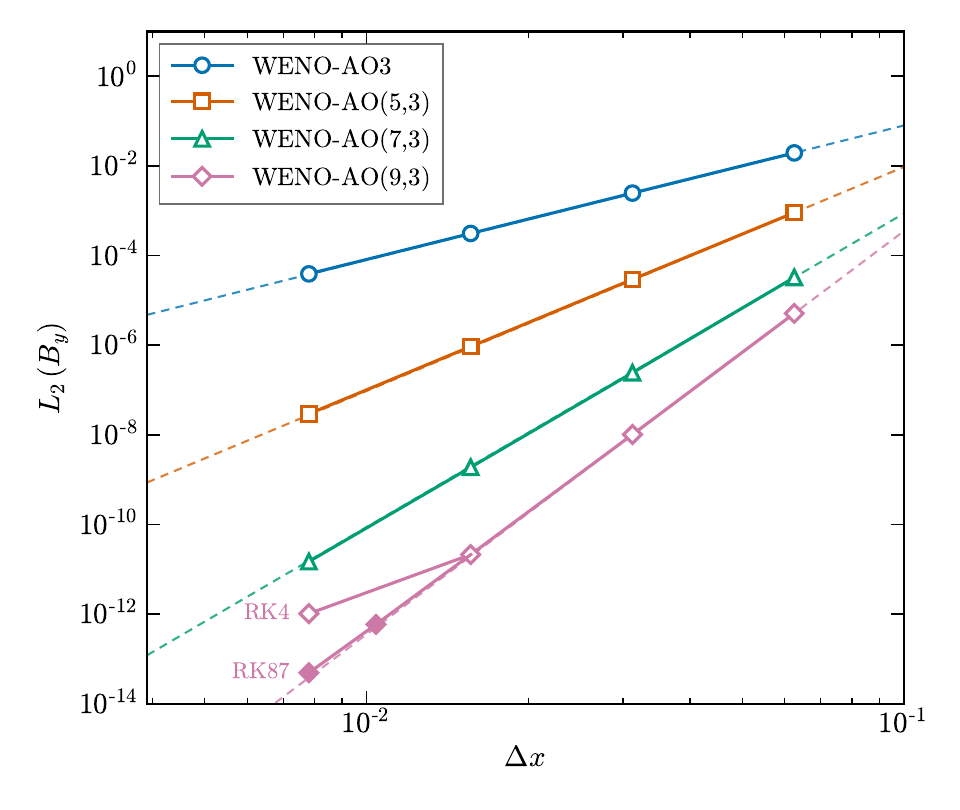}
    \caption{Convergence of the third-, fifth-, seventh-, and ninth-order AFD-WENO-AO schemes for the circularly polarised Alfv\'en wave test at $t=2$. The errors are measured using the $L_2$ norm of $B^y$ relative to the analytic solution. All simulations employ $N_x=16,32,64,$ and $128$, with an additional $N_x=96$ for the ninth-order scheme and $N_y=N_z=8$ throughout. The third-, fifth-, and seventh-order schemes achieve their expected convergence rates, while the ninth-order scheme exhibits ninth-order convergence up to $N_x=64$ points. Beyond this resolution, the convergence rate worsens when using the RK4 time integrator but recovers the expected ninth-order accuracy with RK87.}
    \label{fig:cpalfven}
\end{figure}

Fig.~\ref{fig:cpalfven} presents the $L_2$ errors in $B^y$ relative to the analytic solution at $t=2$. The third-, fifth-, and seventh-order schemes achieve their expected convergence rates, with observed orders of approximately $3$, $5$, and $7$, respectively. The ninth-order scheme also exhibits the expected convergence behaviour between $N_x=16$ and $64$, with observed orders of approximately $9$. At $N_x=128$, however, the error reaches approximately $10^{-12}$ and deviates from the expected ninth-order scaling. Additional calculations with eighth-order Runge--Kutta (RK87) show that the reduction in the order of accuracy at higher resolutions comes from the time integration error. With a more accurate time integrator, the method achieves ninth-order accuracy. Overall, these results confirm that the AFD-WENO-AO schemes achieve their expected spatial convergence orders in smooth test problems, up to ninth-order.

\subsubsection{Balsara Tests}

To assess the ability of the numerical schemes to capture shocks and discontinuities, we perform the five relativistic magnetohydrodynamic shock-tube tests introduced by Balsara~\citep{2001ApJS..132...83B}. These tests cover different physical regimes, including strong magnetic fields, large pressure jumps, and relativistic velocities. We present representative results for Balsara1, Balsara3, and Balsara5, whose initial left and right states are listed in Table~\ref{tab:balsarax}.

\begin{table}
    \centering
    \renewcommand{\arraystretch}{1.3}
    \begin{tabular}{@{}lccc@{}}
        \toprule
        \textbf{Var.} & \textbf{Balsara1} & \textbf{Balsara3} & \textbf{Balsara5} \\
        \midrule
        $\rho$ & 1 \& 0.125 & 1 \& 1 & 1.08 \& 1 \\
        $p$ & 1 \& 0.1 & 1000 \& 0.1 & 0.95 \& 1 \\
        $v_x$ &  0 \& 0 & 0 \& 0 & 0.4 \& -0.45 \\
        $v_y$ &  0 \& 0 & 0 \& 0 & 0.3 \& -0.2 \\
        $v_z$ &  0 \& 0 & 0 \& 0 & 0.2 \& 0.2 \\
        $B_x$ &  0.5 \& 0.5 & 10 \& 10 & 2 \& 2 \\
        $B_y$ &  1 \& -1 & 7 \& 0.7 & 0.3 \& -0.7 \\
        $B_z$ &  0 \& 0 & 7 \& 0.7 & 0.3 \& 0.5 \\
        \bottomrule
    \end{tabular}
    \caption{Initial left and right states for the Balsara1, Balsara3, and Balsara5 tests. The first and second values in each entry correspond to the left and right states, respectively.}
    \label{tab:balsarax}
\end{table}

We evolve these configurations using the third-order AFD-WENO-AO scheme and compare the results with \texttt{GRHydro}~\citep{2014CQGra..31a5005M} and the exact Riemann solver~\citep{2006JFM...562..223G}. Both codes employ an ideal-gas EoS, the HLLE Riemann solver, second-order Runge--Kutta (RK) time integration, and a Courant-Friedrichs-Lewy (CFL) factor of $0.4$, with a uniform grid spacing of $\Delta x=1/1600$. $\Gamma=2$ is used for Balsara1, while $\Gamma=5/3$ is used for the remaining tests. The main difference lies in the spatial discretisation: \texttt{GRHydro} employs the FV method with total variation diminishing (TVD) reconstruction~\citep{tvd_harten_1983} using the monotonised central limiter, whereas \texttt{RAKI} uses third-order AFD-WENO-AO reconstruction and evolves the vector potential. The tests are evolved to $t=0.4$, except for Balsara5, which is evolved to $t=0.55$.

\begin{figure*}
    \centering
    \includegraphics[width=1.0\linewidth]{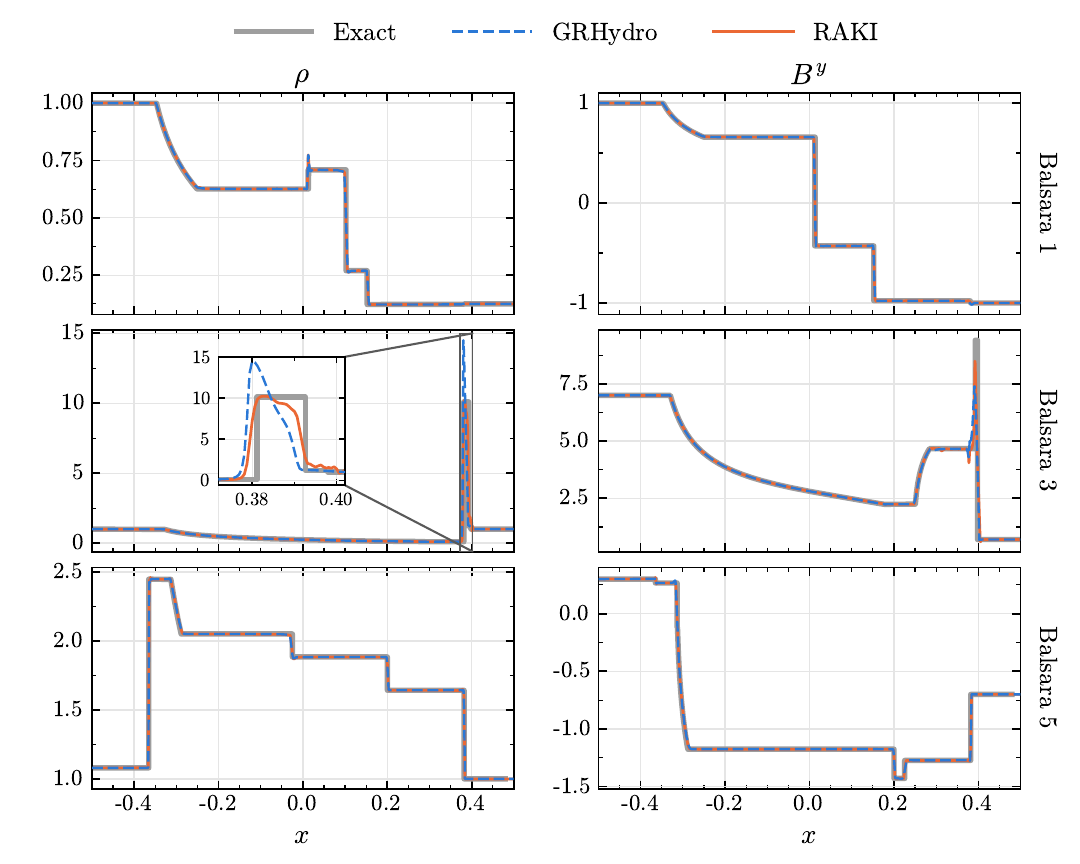}
    \caption{Comparison of the rest-mass density and transverse magnetic field component $B^y$ for the Balsara1, Balsara3, and Balsara5 shock-tube tests. The numerical solutions obtained with third-order AFD-WENO-AO reconstruction in \texttt{RAKI} and TVD reconstruction in \texttt{GRHydro} are compared with the exact Riemann solver. Balsara1 and Balsara3 are shown at $t=0.4$, while Balsara5 is shown at $t=0.55$.}
    \label{fig:balsara}
\end{figure*}

Fig.~\ref{fig:balsara} compares the rest-mass density and the transverse magnetic field component $B^y$. The third-order AFD scheme captures the main shock structures and discontinuities well, showing good agreement with the exact solutions and the results from \texttt{GRHydro} (see the inset of Fig.~\ref{fig:balsara}). This is also reflected in the $L_1$ errors calculated against the exact solutions. For Balsara1, the density error decreases from $2.01\times10^{-3}$ with \texttt{GRHydro} to $1.71\times10^{-3}$ with \texttt{RAKI}, while the error in $B^y$ decreases from $2.84\times10^{-3}$ to $2.55\times10^{-3}$. The largest improvement is seen in Balsara3, where \texttt{RAKI} agrees more closely with the exact solution than \texttt{GRHydro}. In particular, in the thin, dense shell behind the right-moving shock, \texttt{GRHydro} overshoots the exact density peak, while \texttt{RAKI} stays much closer to it. The density error decreases from $8.23\times10^{-2}$ to $3.74\times10^{-2}$, while the error in $B^y$ decreases from $5.15\times10^{-2}$ to $4.62\times10^{-2}$. For Balsara5, both codes give similar results, although \texttt{RAKI} has slightly smaller errors ($2.66\times10^{-3}$ compared to $2.88\times10^{-3}$ in density, and $2.54\times10^{-3}$ compared to $2.93\times10^{-3}$ in $B^y$). Overall, these results show that the AFD-WENO-AO method accurately captures relativistic MHD shock structures and gives results that are comparable to, or better than, those obtained with \texttt{GRHydro}.

We also obtain consistent results for Balsara2, although small oscillations appear near the discontinuities. Balsara4 is also challenging due to its strong initial discontinuity and high magnetic field strength, but our code shows very good agreement with both \texttt{GRHydro} and the exact solution. 

\subsection{2D Tests}\label{sec:2dtest}

The magnetic loop advection test examines how well the numerical scheme preserves the geometry and magnitude of a weak magnetic field as the fluid moves across a two-dimensional domain, while maintaining the divergence-free constraint~\citep{divfree_balsara_2001,divfree_balsara_2009}. We perform this test~\citep{1991JCoPh..92..142D} using the configuration adopted in \texttt{GRHydro}, allowing a direct comparison of the magnetic field evolution. The exact solution consists of a uniform translation of the initial magnetic loop, which returns to its original position after one advection period.

The computational domain is defined by $(x,y)\in[-1,1]\times[-0.5,0.5]$, with periodic boundary conditions in both spatial directions. We set $\rho=1$, $p=3$, $\Gamma=5/3$, and a uniform fluid velocity $(v^x,v^y,v^z)=(1/12,1/24,1/24)$. The initial magnetic loop has a radius $R_{\mathrm{loop}}=0.3$ and an amplitude $A_{\mathrm{loop}}=10^{-3}$. The magnetic field components are defined as
\begin{equation}
    (B^x,B^y)=
    \begin{cases}
        \left(-A_{\mathrm{loop}}\dfrac{y}{r_c},A_{\mathrm{loop}}\dfrac{x}{r_c}\right), & r_c<R_{\mathrm{loop}},\\
        (0,0), & r_c\geq R_{\mathrm{loop}},
    \end{cases}
\end{equation}
where $r_c=\sqrt{x^2+y^2}$ is the radial distance from the centre of the loop. We employ fifth-order AFD-WENO-AO reconstruction for both the fluid and vector-potential evolution, together with second-order RK time integration and a CFL factor of $0.4$.

We first perform three simulations to assess the preservation of the no-monopole constraint: two with uniform grid spacings of $\Delta x=1/16$ and $1/32$, and one using adaptive mesh refinement (AMR), with $\Delta x=1/16$ in the coarse region and $\Delta x=1/32$ in the refined region. We monitor the discrete magnetic divergence throughout the evolution and find that its violation remains at approximately $10^{-16}$ in both the uniform-grid and AMR simulations. This demonstrates that the vector-potential formulation preserves the no-monopole constraint near machine precision, including across refinement boundaries, without employing a staggered grid.

\begin{figure}
    \centering
    \includegraphics[width=1.0\linewidth]{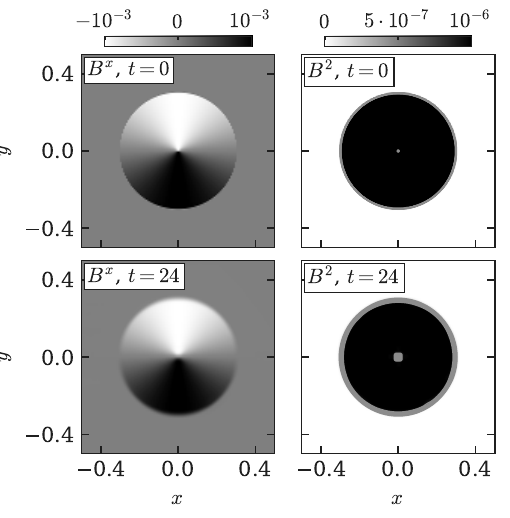}
    \caption{Two-dimensional magnetic loop advection test using fifth-order AFD-WENO-AO reconstruction with $\Delta x=1/256$. The panels show the $x$-component of the magnetic field, $B^x$, and the squared magnetic field strength, $B^2$. The upper row shows the initial configuration at $t=0$, while the lower row shows the configuration at $t=24$, after one complete advection period. The loop largely preserves its geometry and magnetic field strength, with small numerical diffusion concentrated near its boundary.}
    \label{fig:2dloop}
\end{figure}

To assess the accuracy of magnetic field advection, we additionally perform a uniform-grid simulation with $\Delta x=1/256$. Fig.~\ref{fig:2dloop} shows the magnetic field component $B^x$ and the squared magnetic field strength $B^2=B_i B^i$ at the initial time ($t=0$) and after one complete advection period ($t=24$). The magnetic loop maintains its initial circular shape, and the magnetic field strength remains nearly uniform inside the loop. After one period, the plateau value of $B^2$ differs from its initial value by less than $0.5\%$. Numerical diffusion is mainly limited to a thin layer around the initially discontinuous boundary of the loop. The radius at which $B^2$ drops to half of its plateau value decreases from $0.2995$ to $0.2950$. The total magnetic energy also remains close to its initial value, with $\sum B^2(t=24)/\sum B^2(t=0)\approx0.97$. Most of this energy loss occurs during the first unit of time, as the initially sharp boundary becomes smoother. 

\begin{figure}
    \centering
    \includegraphics[width=1.0\linewidth]{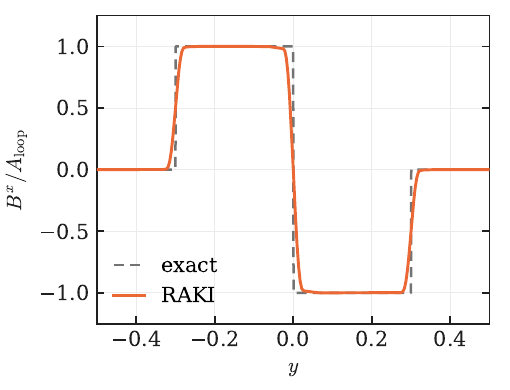}
    \caption{Comparison of the normalised magnetic field component $B^x/A_{\mathrm{loop}}$ along a vertical line through the centre of the magnetic loop with the exact solution at $t=24$. The fifth-order AFD-WENO-AO scheme reproduces the overall magnetic field profile, with small deviations near the discontinuous loop boundaries.}
    \label{fig:loopadvection1d}
\end{figure}

Fig.~\ref{fig:loopadvection1d} further compares the normalised magnetic field component $B^x/A_{\mathrm{loop}}$ along a vertical line through the centre of the loop with the exact solution at $t=24$. The numerical solution closely follows the analytic profile, with the largest deviations occurring near the discontinuous edges of the loop. The measured $L_{1}$ errors in the magnetic field components after one period are approximately $4.1 \times 10^{-6}$, corresponding to about $4.5\%$ of their initial mean absolute values. Compared with the results presented for \texttt{GRHydro}, the magnetic field transitions appear sharper at the same resolution.

\subsection{3D Tests}

\subsubsection{TOV tests}

We perform simulations of stable Tolman--Oppenheimer--Volkoff (TOV)~\citep{tolman_1939, oppenheimer_volkoff_1939} stars to assess the ability of \texttt{RAKI} to maintain hydrostatic equilibrium and evolve magnetic fields in dynamical spacetimes. We adopt the initial stellar configuration used in the \texttt{Einstein Toolkit}, with a polytropic EoS characterised by $\Gamma=2$ and $K=100$, an initial central rest-mass density of $\rho_c=1.28\times10^{-3}$ in geometrised units ($7.90\times10^{14}\,\mathrm{g\,cm^{-3}}$), and a gravitational (baryonic) mass of approximately $1.4\,M_\odot$ ($1.51\,M_\odot$). We consider both unmagnetised and magnetised configurations using fifth-order AFD-WENO-AO reconstruction and fourth-order RK time integration with a CFL factor of $0.25$.

In both cases, we evolve the spacetime using the Baumgarte-Shapiro-Shibata-Nakamura (BSSN) formulation of the Einstein Equations~\citep{bssn_nakamura, shibata_bssn_1995, bssn_baumgarte_1998, bssn_baker}, as implemented in \texttt{McLachlan}~\citep{Brown:2008sb, McLachlan:web, Kranc:web}. The spacetime evolution uses fourth-order finite differences for spatial derivatives. We use four refinement levels, with the finest level having a radius of ${\sim}22\,\mathrm{km}$ and grid spacings of $\Delta x=1477$, $738$, and $369\,\mathrm{m}$ at the three resolutions. These relatively coarse resolutions are chosen to make the numerical errors more visible. 

The computational domain extends to ${\sim}473\,\mathrm{km}$ in each direction, with reflection and $\pi$-rotational symmetries imposed. The atmosphere density is set to $\rho=6.18\times10^{7}\,\mathrm{g\, cm^{-3}}$. We evolve the unmagnetised and magnetised TOV configurations for $5$ and $10\,\mathrm{ms}$, respectively. Since the initial configuration represents a hydrostatic equilibrium solution, the subsequent density oscillations are excited by numerical perturbations introduced during the evolution.

For comparison, the \texttt{GRHydro} results for the same TOV model exhibit maximum central-density deviations of approximately $3.5\%$ and $1.1\%$ at resolutions of $1477$ and $738\,\mathrm{m}$, respectively. At the same resolutions, \texttt{RAKI} produces deviations of approximately $2.9\%$ and $0.82\%$. Although the initial stellar configurations are identical, the numerical methods differ. In particular, \texttt{GRHydro} employs the Piecewise Parabolic Method (PPM)~\citep{ppm_colella_1984} for reconstruction and the Marquina Riemann solver~\citep{marquina_1996}, whereas \texttt{RAKI} uses fifth-order AFD-WENO-AO reconstruction and the HLLE Riemann solver. These results show that \texttt{RAKI} produces smaller maximum density deviations than \texttt{GRHydro} at the same resolutions.

\begin{figure}
    \centering
    \includegraphics[width=1.0\linewidth]{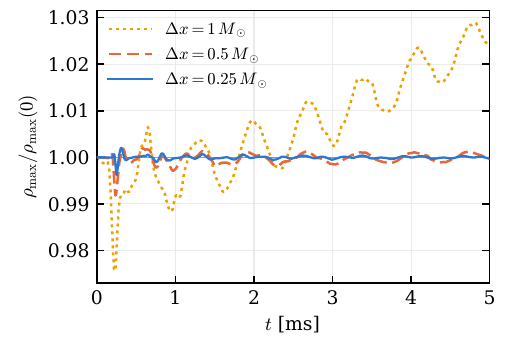}
    \caption{Relative evolution of the maximum rest-mass density of an unmagnetised TOV star using fifth-order AFD-WENO-AO reconstruction. The three curves correspond to finest grid spacings of $\Delta x= \{1, 0.5, 0.25\}\,M_\odot$ ($\{1477, 738, 369\}\, \mathrm{m}$). The amplitude of the density oscillations decreases with increasing resolution, demonstrating improved preservation of hydrostatic equilibrium.}
    \label{fig:tovoscillation}
\end{figure}

Fig.~\ref{fig:tovoscillation} shows the evolution of the normalised maximum rest-mass density, $\rho_{\max}(t)/\rho_{\max}(0)$, at the three resolutions. The amplitude of the oscillations decreases systematically with increasing resolution. At the lowest resolution, the density deviations range from approximately $-2.43\%$ to $+2.89\%$, decreasing to $-0.82\%$ and $+0.19\%$ at the intermediate resolution and $-0.38\%$ and $+0.20\%$ at the highest resolution.

We also examine the radial oscillation frequencies excited by numerical perturbations. The fundamental radial-mode frequency is approximately $1.45\,\mathrm{kHz}$ at all three resolutions, in agreement with the reference value of $1.442\,\mathrm{kHz}$ reported by~\cite{2002PhRvD..65h4024F}. This result indicates that \texttt{RAKI} reproduces the expected fundamental radial mode frequency while maintaining the equilibrium configuration.

For the magnetised TOV test, we initialise a purely poloidal magnetic field with a maximum initial field strength of ${\sim}10^{15}\,\mathrm{G}$, similar to the field strengths used in~\cite{2014CQGra..31a5005M, 2015CQGra..32q5009E}, using the vector potential in Eq.~\eqref{eq:aphi}. We use the generalised Lorenz gauge with a damping parameter $\xi=1.5/\Delta t_{\max}$.

\begin{figure*}[t]
    \centering
    \includegraphics[width=\textwidth]{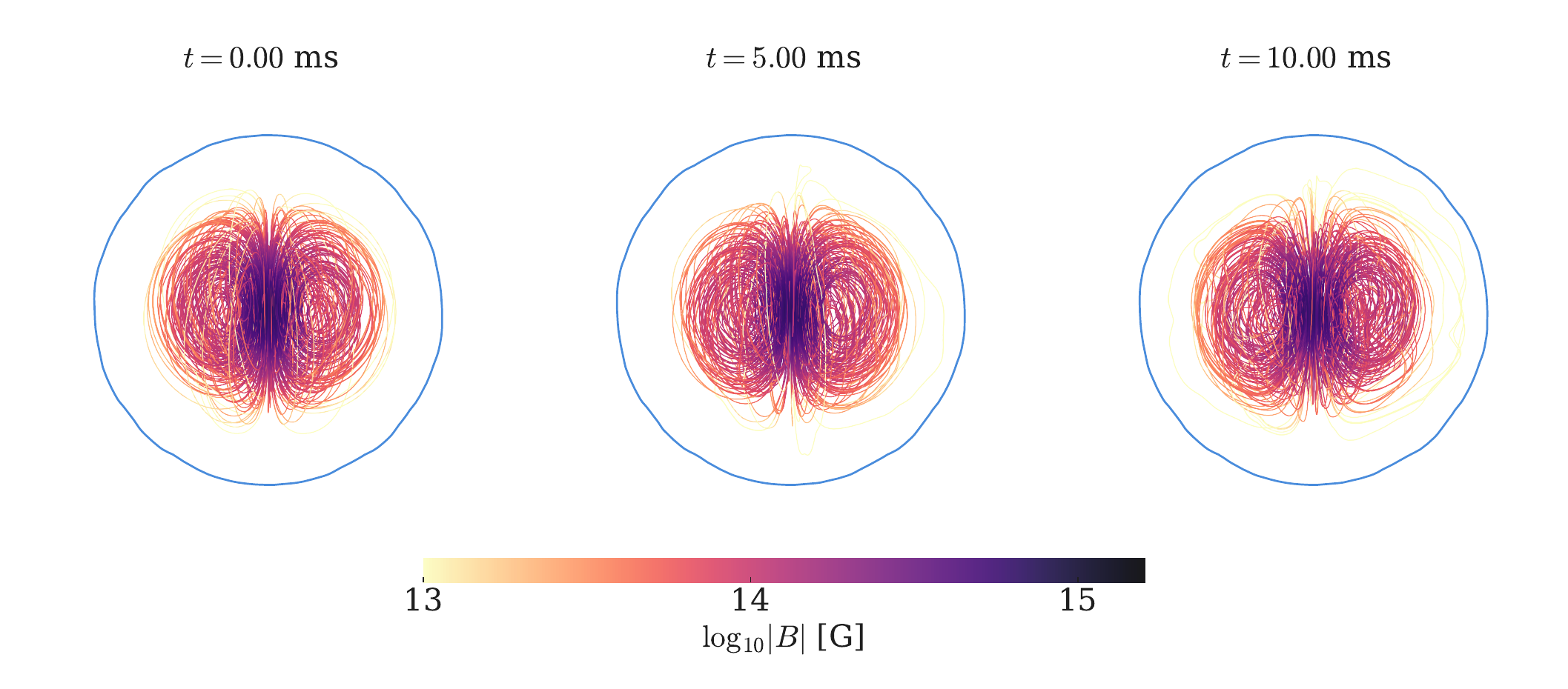}
    \caption{Evolution of the magnetic field in a magnetised TOV star at $t=0.00$, $5.00$, and $10.00\,\mathrm{ms}$. The panels show three-dimensional magnetic field lines coloured by the magnetic field strength, $|B|$, in Gauss. The initial magnetic field has a maximum strength of approximately $10^{15}\,\mathrm{G}$ and is confined to the stellar interior. During the evolution, the initially ordered poloidal field becomes distorted. The blue contours mark a rest-mass density of $\rho=10^{13}\,\mathrm{g\,cm^{-3}}$.}
    \label{fig:magtov}
\end{figure*}

Fig.~\ref{fig:magtov} shows the three-dimensional structure of the initially poloidal magnetic field at three representative times. The magnetic field remains confined within the stellar interior throughout the evolution. At later times, internal fluid motions deform the field geometry slightly, bending and stretching the initially poloidal loops. 

\begin{figure}
    \centering
    \includegraphics[width=1.0\linewidth]{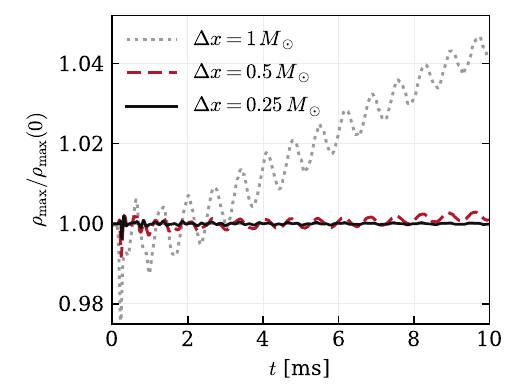}
    \caption{Relative evolution of the maximum rest-mass density of a magnetised TOV star using fifth-order AFD-WENO-AO reconstruction. The three curves correspond to finest grid spacings of $\Delta x=1477$, $738$, and $369\,\mathrm{m}$. The amplitude of the density oscillations decreases with increasing resolution, demonstrating improved preservation of hydrostatic equilibrium.}
    \label{fig:magtovrho}
\end{figure}

\begin{figure}
    \centering
    \includegraphics[width=1.0\linewidth]{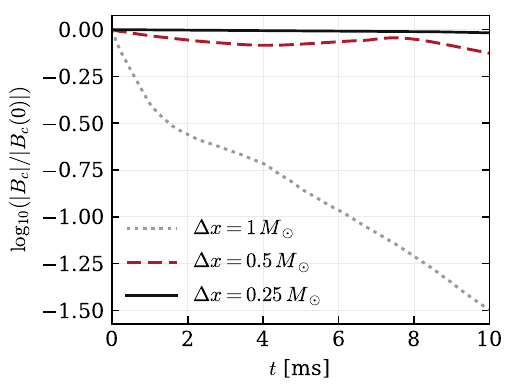}
    \caption{Evolution of the central magnetic field strength normalised to its initial value, shown on a logarithmic scale, for a TOV star using fifth-order AFD-WENO-AO reconstruction. The three curves correspond to finest grid spacings of $\Delta x=1477$, $738$, and $369\,\mathrm{m}$. The magnetic field shows a resolution-dependent decrease, with significantly better preservation at higher resolutions.}
    \label{fig:magtovB}
\end{figure}

Figs.~\ref{fig:magtovrho} and~\ref{fig:magtovB} show that the maximum rest-mass density closely follows the corresponding unmagnetised simulation, while the magnetic field is better preserved at higher resolutions. At the highest resolution, the central magnetic field retains approximately $96.1\%$ of its initial value by the end of the evolution. The radial oscillation frequencies also remain nearly unchanged compared to the unmagnetised case. Furthermore, the discrete magnetic divergence is maintained at round-off levels throughout the simulation, with $\|\nabla\cdot\mathbf{B}\|_\infty$ of order $10^{-20}$. These results demonstrate that the non-staggered, node-centred vector-potential formulation can evolve magnetised neutron stars while preserving the no-monopole constraint to machine precision.

\begin{figure}
    \centering
    \includegraphics[width=1.0\linewidth]{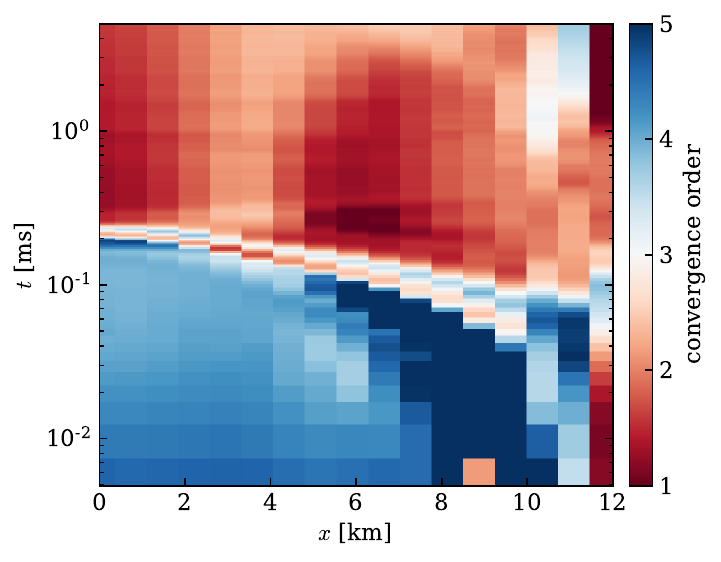}
    \caption{Convergence order of the rest-mass density along the $x$-axis of the static TOV star, calculated using resolutions $\Delta x=0.5$ and $0.25\,M_\odot$. The colour scale is capped at the formal fifth-order accuracy of the scheme. The convergence order decreases near the stellar surface and progressively propagates towards the centre at later times, suggesting that errors associated with the atmosphere treatment propagate into the stellar interior.}
    \label{fig:convtov}
\end{figure}

Fig.~\ref{fig:convtov} shows the convergence order of the rest-mass density in the unmagnetised TOV simulation. The convergence order is close to the expected fifth-order in the stellar interior but decreases near the surface and in the atmosphere. At later times, this reduction in accuracy extends towards the stellar centre, suggesting that errors associated with the atmosphere treatment propagate into the interior. Improving the atmosphere treatment may therefore be necessary to maintain high-order convergence over longer evolutions.

\subsubsection{Binary neutron star merger}

Having completed the one-, two-, and three-dimensional tests to assess the accuracy and robustness of our code, we now consider a more challenging application: binary neutron star merger simulation. We perform magnetised and unmagnetised simulations of an equal-mass, irrotational binary neutron star (BNS) merger using the LORENE~\citep{lorene_2001, lorene_2016} initial data available in the \texttt{Einstein Toolkit}. Each star has a baryonic mass of $1.445\,M_\odot$ and a gravitational mass of $1.317\,M_\odot$, with an initial coordinate separation of $45\,\mathrm{km}$. We use a polytropic EoS with $\Gamma=2$ and $K=123.6$, together with a thermal component $\Gamma_{\mathrm{th}}=2$ to account for shock heating during merger.

We employ fifth-order AFD-WENO-AO reconstruction for both the fluid and vector potential evolution. For the magnetised models, we initialise a poloidal magnetic field with a maximum strength of $10^{15}\,\mathrm{G}$. The spacetime is evolved using the Conformal and Covariant Z4 \texttt{(CCZ4)} formulation of the Einstein Equations~\citep{Alic_ccz4, Alic_2013} implemented in \texttt{McLachlan}~\citep{Brown:2008sb, McLachlan:web, Kranc:web}, with sixth-order finite differences in both unmagnetised and magnetised simulations to reduce errors arising from the spacetime discretisation. Time evolution is performed using RK4 with a CFL factor of $0.25$. The computational domain extends to ${\sim}851\,\mathrm{km}$ in each direction, with reflection and $\pi$-rotational symmetries imposed, as in the TOV tests. Six refinement levels are used, the finest of which covers each star within a radius of ${\sim}19.2\,\mathrm{km}$ during inspiral and has grid spacings (in the three cases used for the convergence study below) of $\Delta x=554$, $369$, and $277\,\mathrm{m}$. The atmosphere density is set to $\rho=6.18\times10^{6}\,\mathrm{g\,cm^{-3}}$.

\begin{figure*}
    \centering
    \includegraphics[width=1.0\linewidth]{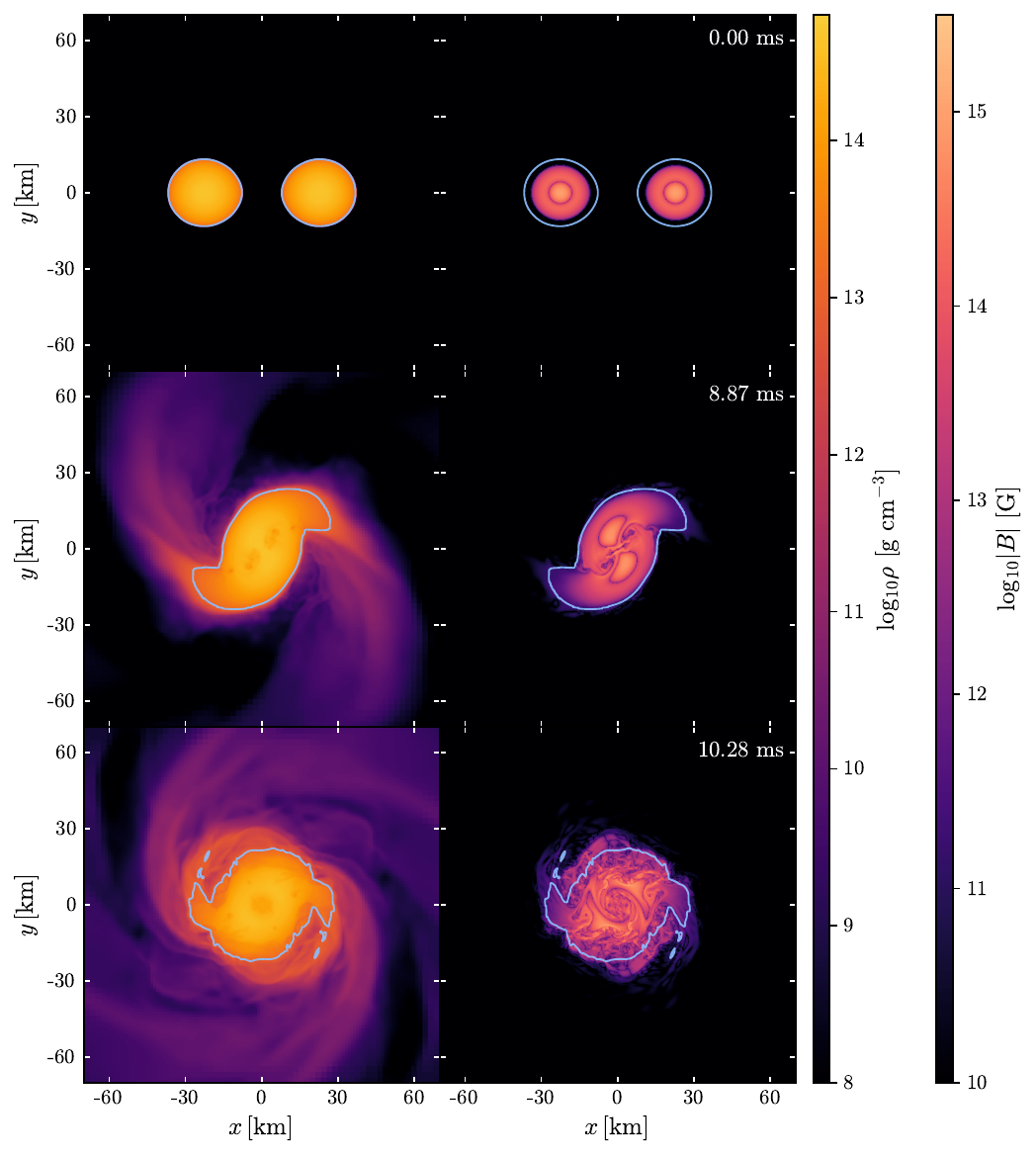}
    \caption{Snapshots of the rest-mass density (left column) and magnetic field strength (right column) in the equatorial plane during the binary neutron star merger. From top to bottom, the panels correspond to $t=0.00$, $8.87$, and $10.28\,\mathrm{ms}$ for the highest resolution simulation ($\Delta x = 277\, \mathrm{m}$). The contours mark a rest-mass density of $\rho=10^{13}\,\mathrm{g\,cm^{-3}}$.}
    \label{fig:2dimages}
\end{figure*}

Fig.~\ref{fig:2dimages} shows snapshots of the rest-mass density and magnetic field in the equatorial plane at three representative times. During inspiral, the two neutron stars approach each other while maintaining their individual high-density cores. At contact, strong shear develops at the interface between the stars, where the Kelvin--Helmholtz instability can amplify the magnetic field. Following merger, the matter forms a rapidly rotating remnant with a more complex magnetic field structure. These snapshots demonstrate that \texttt{RAKI} can evolve the fluid and magnetic field through the merger and into the early post-merger phase using the same high-order AFD-WENO-AO framework.

The BNS merger simulation also provides a test of the no-monopole constraint in a dynamical spacetime. As discussed in Sec.~\ref{sec:2dtest}, our two-dimensional AMR tests demonstrate that the no-monopole constraint can be preserved near machine precision without employing a staggered grid. BNS mergers provide a more demanding test because the refinement levels follow the stars during inspiral. Despite these additional complications, \texttt{RAKI} maintains the discrete magnetic divergence below $10^{-16}$. This result demonstrates that the non-staggered, node-centred vector-potential formulation preserves the no-monopole constraint at machine precision even in a dynamical spacetime.

We next examine the convergence of the gravitational-waves for both unmagnetised and magnetised models. The gravitational-wave strain is calculated by fixed-frequency integration of the Newman-Penrose scalar $\Psi_4$~\citep{ffi_reisswig_2011}, extracted at a radius of ${\sim}148\,\mathrm{km}$. We compare the phases of the dominant $(l,m)=(2,2)$ mode at three resolutions to estimate the accumulated phase error up to merger.

\begin{figure*}
    \centering
    \includegraphics[width=1.0\linewidth]{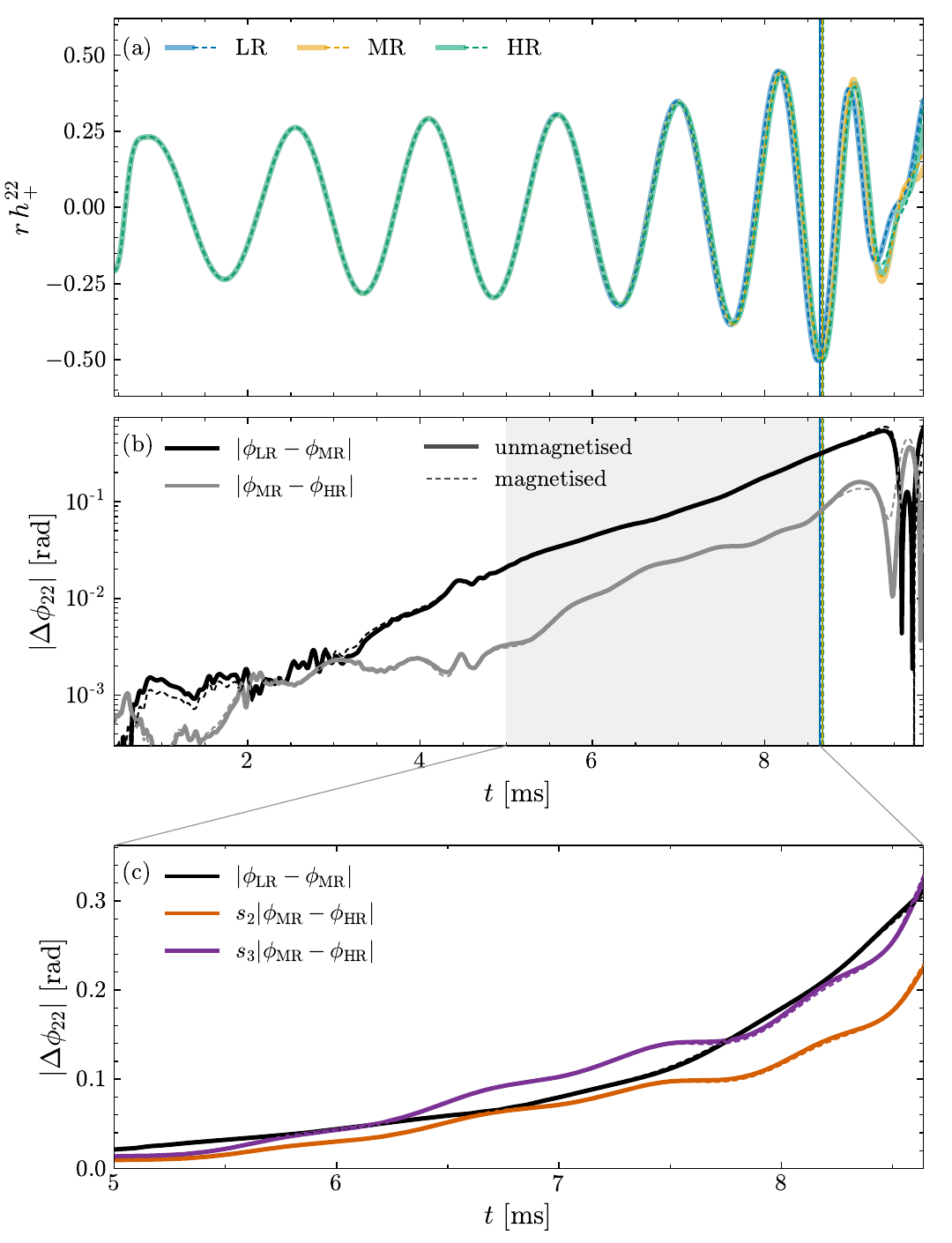}
    \caption{Gravitational-wave strain and phase convergence for unmagnetised and magnetised binary neutron star merger simulations. The upper panel shows the waveforms at three resolutions for both models, while the middle panel compares the phase differences between successive resolutions. The lower panel provides a magnified view of the phase from $5\,\mathrm{ms}$ up to merger on a linear scale. The rescaled phase differences are used to assess the observed convergence order during inspiral.}
    \label{fig:gwconvergencehydro}
\end{figure*}

Fig.~\ref{fig:gwconvergencehydro} presents the GW strain and phase differences for both unmagnetised and magnetised BNS mergers. The upper panel shows the GW strain at three resolutions for each model, while the middle panel compares the phase differences between successive resolutions. The lower panel compares the rescaled phase differences from $5\,\mathrm{ms}$ up to merger on a linear scale. The waveforms remain in close agreement during the early inspiral, with increasing phase differences towards merger. The convergence behaviour provides a test of the accuracy of the numerical evolution in a dynamical spacetime, complementing the formal high-order convergence demonstrated in the smooth test problems.

The middle and lower panels show approximately third-order convergence of the GW phase in both unmagnetised and magnetised simulations. Consistent with the TOV results, the reduced convergence order relative to the formal order of the numerical scheme is likely associated with the treatment of the stellar surface and atmosphere.

\subsection{Performance}

In this subsection, we compare the performance of our new code, \texttt{RAKI}, with \texttt{IllinoisGRMHD}~\cite{2015CQGra..32q5009E}, a well-established and widely used code for GRMHD simulations. The main reason for comparing the performance with \texttt{IllinoisGRMHD} is that it also employs the vector potential approach. To ensure a fair comparison, we keep all parameters identical, except for those that differ due to the numerical implementations of the two codes. We perform magnetised TOV simulations using WENO-AO(5,3) and WENO-AO3 reconstruction in \texttt{RAKI} and PPM reconstruction in \texttt{IllinoisGRMHD}. For the curl calculation, \texttt{RAKI} employs high-order central differencing, while \texttt{IllinoisGRMHD} uses second-order staggered differencing. In both codes, the metric derivatives appearing in the source terms are computed using fourth-order finite differences, and the same $\Gamma$-law EoS is adopted. All three simulations use the same finest-grid resolution of $369\,\mathrm{m}$ and a CFL factor of $0.25$. Although \texttt{RAKI} supports both rotational and reflection symmetries, these symmetries are not used in any of the three simulations. Thus, the main differences between the two codes lie in the underlying numerical schemes (AFD versus FV), the reconstruction methods (WENO-AO(5,3) and WENO-AO3 versus PPM), and the discretisation of the curl operator.

\begin{figure}
    \centering
    \includegraphics[width=1.0\linewidth]{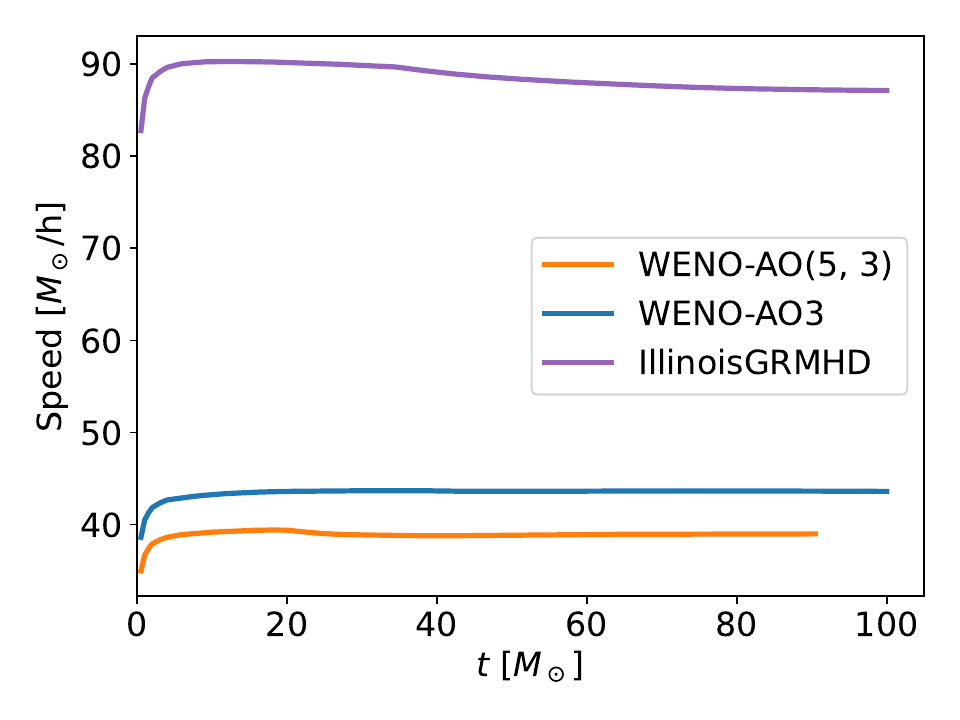}
    \caption{Code speed in the magnetised TOV tests for \texttt{RAKI} using AFD-WENO-AO(5,3) and AFD-WENO-AO3 reconstruction, compared with \texttt{IllinoisGRMHD}. The evolution time per wall-clock hour is $39\,M_\odot$ and $46.5\,M_\odot$ for the two \texttt{RAKI} schemes, respectively, compared with $86.5\,M_\odot$ for \texttt{IllinoisGRMHD}. Our code \texttt{RAKI} is approximately two times slower than \texttt{IllinoisGRMHD}; however, it shows up to ninth-order accuracy in smooth regions and also demonstrates third-order convergence in gravitational-wave phase.}
    \label{fig:igmcomp}
\end{figure}

Fig.~\ref{fig:igmcomp} shows the performance differences between \texttt{IllinoisGRMHD} and \texttt{RAKI} with two different reconstruction schemes. As seen in the figure, \texttt{RAKI} is approximately two times slower than \texttt{IllinoisGRMHD}. However, for the same physical setup, a magnetised TOV star evolved up to $t \approx 90\,M_\odot$ at the finest resolution of $\Delta x = 0.25\,M_\odot$, the AFD-WENO-AO(5,3) scheme performs better than \texttt{IllinoisGRMHD}, particularly in preserving the equilibrium structure of the star and its magnetic field.

Between $t = 60\,M_\odot$ and $t = 90\,M_\odot$, the average deviation of the central density is only $+0.003\%$ for AFD-WENO-AO(5,3), compared to $-0.14\%$ for \texttt{IllinoisGRMHD}. This indicates that the average stellar structure remains nearly unchanged with AFD-WENO-AO(5,3), while \texttt{IllinoisGRMHD} shows a gradual drift from the initial equilibrium. The magnetic field is also better preserved with AFD-WENO-AO(5,3). At $t = 90\,M_\odot$, the maximum magnetic field strength decreases by $0.2\%$ of its initial value in \texttt{RAKI}, compared with $9.38\%$ in \texttt{IllinoisGRMHD}. Similarly, the average decrease in the $L_2$ norm of the magnetic field is only $0.01\%$ for AFD-WENO-AO(5,3), compared to $0.21\%$ for \texttt{IllinoisGRMHD}.

Additional \texttt{IllinoisGRMHD} simulations show that, over the first $2\,\mathrm{ms}$, the maximum magnetic field strength decreases by $9.4\%$ and $1.6\%$ at $\Delta x\approx369\,\mathrm{m}$ and $148\,\mathrm{m}$, respectively, compared with only $0.8\%$ for \texttt{RAKI} at $369\,\mathrm{m}$. The measured effective convergence order ($p\approx2$) suggests that \texttt{IllinoisGRMHD} would require a grid spacing of approximately $100\,\mathrm{m}$ to achieve comparable magnetic field preservation. Assuming a computational cost scaling of $\Delta x^{-4}$ and accounting for \texttt{IllinoisGRMHD} being approximately $2.2$ times faster at fixed resolution, this corresponds to nearly two orders of magnitude higher computational cost. This estimate is based on extrapolation beyond the tested resolutions and is specific to the TOV test.

Overall, these results show that although \texttt{RAKI} is computationally more expensive than \texttt{IllinoisGRMHD}, the AFD-WENO-AO(5,3) scheme provides better preservation of the stellar equilibrium and magnetic field. The smaller deviations in central density and magnetic field strength suggest that AFD-WENO-AO(5,3) introduces less numerical dissipation. Thus, the additional computational cost of \texttt{RAKI} is more than compensated for by its improved accuracy in preserving the stellar structure and magnetic field over time.

\section{Summary and Conclusion}\label{sec:conclusion}

In this work, we have presented \texttt{RAKI}, a new fully general relativistic ideal magnetohydrodynamics code within the \texttt{Einstein Toolkit} framework. We employ the alternative finite difference (AFD) method~\citep{balsaraAlternativeFiniteDifference2025a} for both the fluid and magnetic field evolution. In combination with the AFD approach, we use adaptive-order WENO interpolation to obtain highly accurate numerical solutions while maintaining robustness. One of the main advantages of this code is its node-centred approach, which avoids the complexity of staggered-grid structure.

We validate the code using the circularly polarised Alfv\'en-wave test, for which we observe ninth-order convergence. The Balsara shock tube tests further demonstrate the code's ability to accurately capture the behaviour of strong shocks. We further demonstrate the robustness of the implementation by performing $2$-D loop advection, magnetised TOV and binary neutron star merger simulations. 

The magnetic-loop advection, magnetised TOV, and BNS merger simulations demonstrate that the non-staggered vector-potential formulation preserves the discrete no-monopole constraint to machine precision, even during BNS mergers with moving refinement regions. This is particularly important because staggered grids increase code complexity and can make the implementation of additional physics, such as resistivity, more challenging.

The TOV tests also reveal the limitations of high-order accuracy near stellar surfaces. The observed convergence order approaches the expected fifth-order in the stellar interior but decreases near the surface and atmosphere. At later times, this reduction extends towards the stellar centre, suggesting that the atmosphere treatment limits the accuracy of longer evolutions. Improving this treatment is therefore an important direction for further development.

We compare the computational performance and accuracy of \texttt{RAKI} with \texttt{IllinoisGRMHD} using magnetised TOV simulations. Although \texttt{RAKI} is approximately twice as slow at a fixed grid spacing of ${\sim}369\,\mathrm{m}$, it provides significantly better preservation of the stellar equilibrium and magnetic field. Over the first $2\,\mathrm{ms}$, the maximum magnetic field strength decreases by only $0.8\%$ in \texttt{RAKI}, compared with $9.4\%$ in \texttt{IllinoisGRMHD} at the same resolution. Extrapolating the measured resolution dependence of \texttt{IllinoisGRMHD} suggests that achieving comparable magnetic field preservation would require approximately $100\,\mathrm{m}$ grid spacing, corresponding to nearly two orders of magnitude greater computational cost under the assumed cost scaling. Although specific to the TOV test, this comparison suggests that the higher computational cost per resolution of \texttt{RAKI} can be offset by substantially improved magnetic field preservation.

We also obtain approximately third-order convergence of the gravitational-wave phase in both magnetised and unmagnetised BNS merger simulations using fifth-order AFD-WENO-AO reconstruction in \texttt{RAKI}. To the best of our knowledge, together with its computational performance, these results suggest that \texttt{RAKI} is the most efficient CPU-based implementation for the class of simulations considered here.

In future work, we will perform BNS merger simulations with finite-temperature, composition-dependent equations of state and different magnetic field topologies, incorporating finite electrical resistivity, and port \texttt{RAKI} to GPUs to improve computational efficiency.

\section*{Acknowledgements}
We would like to thank Jonathan Thompson for valuable discussions. This work used the DiRAC Memory Intensive service (Cosma8) at Durham University via RAC18 allocation, managed by the Institute for Computational Cosmology on behalf of the STFC DiRAC HPC Facility (www.dirac.ac.uk). The DiRAC service at Durham was funded by BEIS, UKRI and STFC capital funding, Durham University and STFC operations grants. DiRAC is part of the UKRI Digital Research Infrastructure. RM also acknowledges the use of the IRIDIS High Performance Computing Facility, and associated support services at the University of Southampton, in the completion of this work. NA and IH also gratefully acknowledge support from the Science and Technology Facility Council (STFC) via grant number ST/Y000811/1.

\section*{Use of Large Language Models}
We acknowledge the use of Claude (Anthropic), a large language model, to assist with code refactoring, debugging, and validation during the final stages of code development. All suggested modifications were reviewed and tested by the authors.

\appendix

\section{AFD-WENO details}
\label{sec:AFD_details}

We here briefly summarise the steps outlined in~\cite{balsaraEfficientAlternativeFinite2024} for constructing the scheme. There are three steps to constructing an AFD-WENO scheme.

\subsection{Update formula}
\label{sec:AFD_details_update}

Given a system of PDEs
\begin{equation}
    \pdv{\mathbf{q}}{t} + \pdv{\mathbf{f}(\mathbf{q})}{x} = 0 \, ,
\end{equation}
the aim is to construct a high-order accurate update formula
\begin{equation}
    \odv{}{t} \mathbf{q}_i = \frac{1}{\Delta x} \left[ \hat{\mathbf{f}}_{i-1/2} - \hat{\mathbf{f}}_{i+1/2} \right] \, .
\end{equation}

The numerical flux will be written as
\begin{equation}
    \label{eq:AFD_flux2}
    \hat{\mathbf{f}}_{i+1/2} = \mathbf{f}^*\left( \mathbf{q}^-_{i+1/2}, \mathbf{q}^+_{i+1/2} \right) + \sum c_{2k} \left( \Delta x \right)^{2 k} \left[ \partial_x^{2 k} \mathbf{f} \right]_{i+1/2} \, .
\end{equation}
Here $\mathbf{q}^\pm_{i+1/2}$ is the data reconstructed to the ``cell interface'' location and $\mathbf{f}^*(\mathbf{q}_L, \mathbf{q}_R)$ is the (possibly approximate) solution to the Riemann problem given the reconstructed data.

A Taylor series expansion of the flux about the vertex $x_i$ can be used to construct the coefficients $c_{2k}$. The coefficients necessary for a seventh-order scheme are
\begin{equation}
    c_2 = -\frac{1}{24} \, , \quad c_4 = \frac{7}{\num{5760}} \, , \quad c_6 = -\frac{31}{\num{967680}} \, .
\end{equation}

To complete the calculation of the correction terms in~\zcref{eq:AFD_flux1} we need high-order spatial derivatives of the flux, evaluated at the cell interface locations. It is more computationally practical to manipulate the equation using the Jacobian characteristic matrix to write
\begin{equation}
    \label{eq:AFD_flux3}
    \begin{aligned}
    \hat{\mathbf{f}}_{i+1/2} &= \mathbf{f}^*\left( \mathbf{q}^-_{i+1/2}, \mathbf{q}^+_{i+1/2} \right) \\ &\quad+ \sum c_{2k} \left( \Delta x \right)^{2 k}  \left[ \partial_x^{2 k-1} \left( \pdv{\mathbf{f}}{\mathbf{q}} \partial_x \mathbf{q}  \right)\right]_{i+1/2} \, .
    \end{aligned}
\end{equation}
The matrix $\mathbf{J} = \partial_{\mathbf{q}} \mathbf{f}$ can be evaluated algebraically at node centres. Therefore the term $\mathbf{J} \partial_x \mathbf{q}$ can be efficiently computed at nodes, using the same reconstruction as needed to solve the Riemann problem in the computation of $\partial_x \mathbf{q}$.

\subsection{Reconstruction of the fields}
\label{sec:AFD_details_recon}

The vertex-centred data $\mathbf{q}_i$ needs reconstructing over the cell as $\mathbf{P}(x)$ for two purposes. First, the reconstruction is evaluated at the cell interfaces to give the data $\mathbf{q}^+_{i-1/2},\mathbf{q}^-_{i+1/2}$ which will be needed for the Riemann problem solutions at the left and right interfaces of the zone. Second, the derivative $\partial_x \mathbf{P}(x)$ is evaluated at the vertex in order to construct the term $\partial_{\mathbf{q}} \mathbf{f} \, \partial_x \mathbf{q}$ at the vertex. This will later be used in the construction of the correction terms in~\zcref{eq:AFD_flux3}.

In the remainder of this subsection and the next we will consider applying the reconstruction to a single scalar variable. The notation will reflect this. In extending to a system, the reconstruction can either be applied component-wise (simple, but risks introducing oscillations) or characteristic-wise (more complex and expensive, especially for relativistic MHD where the characteristic decomposition has a range of degeneracies, but minimizes oscillations). The computationally optimal strategies are discussed in detail in~\cite{balsaraEfficientFiniteDifference2024}. For simplicity, in this paper we use component-wise reconstruction.

The AFD-WENO method of~\cite{balsaraEfficientFiniteDifference2024} relies on a standard third-order centred WENO finite difference scheme. Higher-order methods are constructed, and are then hybridised with the third-order scheme (in short, they fall back to the third-order method if the high-order results are too oscillatory).

In these WENO schemes the final stencil is given as a convex combination of simple stencils $P_k(x)$ centred on particular vertices,
\begin{equation}
    \label{eq:AFD_stencil}
    P(x) = \sum \bar{\omega}_k P_k(x) \, , \quad \sum \bar{\omega}_k = 1 \, .
\end{equation}
The un-normalized weights $\omega_k$ are given as a combination of the linear weights $\gamma_k$, the smoothness indicators for each individual stencil $\beta_k$, the ``global'' smoothness indicator $\tau$, and a small number $\epsilon$ to avoid division by zero, as
\begin{equation}
    \label{eq:AFD_weights_unnorm}
    \omega_k = \gamma_k \left( 1 +  \frac{\tau^2}{\left( \beta_k + \epsilon \right)^2} \right) \, .
\end{equation}
The scheme is fixed by the choice of linear weights $\gamma_k$ (typically linked to the optimal weights, so the order of the scheme is as high as possible), and the smoothness indicators $\beta_k$ (typically linked to higher derivatives of the associated stencil $P_k$) and $\tau$ (typically a combination of the local indicators). Once the un-normalised weights are known, the final weights are computed as
\begin{equation}
    \label{eq:AFD_weights_norm}
    \bar{\omega}_k = \frac{\omega_k}{\sum_l \omega_l} \, .
\end{equation}

For the AFD-WENO method of~\cite{balsaraEfficientAlternativeFinite2024} the stencils $P_k(x)$ are expressed as a series of Legendre polynomials $L_\ell(x)$ as
\begin{equation}
    \label{eq:AFD_stencils}
    P_k(x) = \sum_{\ell=0} u^{(\ell)}_k L_\ell(x) \, ,
\end{equation}
where the Legendre polynomials are scaled to span the interval $[-1/2, 1/2]$ and the highest order polynomial term is scaled to $1$. The $u^{(\ell)}_k$ are essentially the interpolation coefficients for the stencils. Assuming that the central point of interest is at $x_0$ and is denoted $q_0$, the three stencils used in the third order scheme therefore use the Legendre polynomials
\begin{equation}
    \label{eq:AFD_k3_Lp}
    L_0(x) = 1 \, \quad L_1(x) = x \, \quad L_2(x) = x^2 - \tfrac{1}{12} \, .
\end{equation}
The interpolation coefficients determining the stencils are therefore given by
\begin{widetext}
\begin{equation}
    \label{eq:AFD_k3_stencils}
    \begin{aligned}
        u^{(0)}_1 &= \tfrac{1}{24} \left( 25 q_0 - 2 q_{-1} + q_{-2} \right) & u^{(1)}_1 &= \tfrac{1}{2} \left( 3 q_0 - 4 q_{-1} + q_{-2} \right) & u^{(2)}_1 &= \tfrac{1}{2} \left( q_0 - 2 q_{-1} + q_{-2} \right) \, , \\
        u^{(0)}_2 &= \tfrac{1}{24} \left( 22 q_0 + q_{-1} + q_{1} \right) & u^{(1)}_2 &= \tfrac{1}{2} \left( q_1 - q_{-1} \right) & u^{(2)}_2 &= \tfrac{1}{2} \left( -2 q_0 + q_{-1} + q_{1} \right) \, , \\
        u^{(0)}_3 &= \tfrac{1}{24} \left( 25 q_0 - 2 q_{1} + q_{2} \right) & u^{(1)}_3 &= \tfrac{1}{2} \left( -3 q_0 + 4 q_{1} - q_{2} \right) & u^{(2)}_3 &= \tfrac{1}{2} \left( q_0 - 2 q_{1} + q_{2} \right) \, .
    \end{aligned}
\end{equation}
\end{widetext}
We see that the coefficients $u^{(\ell)}_1$, associated with the stencil $P_1$, are left-biased (with respect to the central point $q_0$), and similarly the stencil $P_2$ is centred and the stencil $P_3$ is right-biased.

The linear weights are chosen with respect to a single parameter $\gamma_\mathrm{Lo} \in [0.8, 0.95]$ that fixes how heavily the centred stencil is weighted. The individual linear weights are then
\begin{equation}
    \label{eq:AFD_linear_weights}
    \gamma_1 = \tfrac{1}{2} \left( 1 - \gamma_\mathrm{Lo} \right) \, , \quad \gamma_2 = \gamma_\mathrm{Lo} \, \quad \gamma_3 = \gamma_1 \, .
\end{equation}

The smoothness indicators are constructed from the integrated squares of derivatives of the interpolation polynomials. For the third-order case, they are given by
\begin{equation}
    \label{eq:AFD_k3_smoothness1}
    \beta_k = \left( u^{(1)}_k \right)^2 + \frac{13}{3} \left( u^{(2)}_k \right)^2 \, 
\end{equation}
for the local smoothness indicators, and
\begin{equation}
    \label{eq:AFD_k3_smoothness2}
    \tau = \frac{1}{2} \left( \left| \beta_2 - \beta_1 \right| + \left| \beta_3 - \beta_2 \right| \right) 
\end{equation}
for the global indicator.

The practical implementation of the third order scheme therefore starts from the five points $q_i$ with $i = -2, \dots, 2$. The interpolation coefficients in~\zcref{eq:AFD_k3_stencils} are then computed, giving the individual stencils from~\zcref{eq:AFD_stencils}. Then the various smoothness indicators from~\zcref{eq:AFD_k3_smoothness1,eq:AFD_k3_smoothness2} are computed which, combined with the linear weights from~\zcref{eq:AFD_linear_weights} gives the stencil weights $\bar{\omega}_k$ from~\zcref{eq:AFD_weights_unnorm} and~\zcref{eq:AFD_weights_norm}. This gives the piecewise polynomial WENO stencil $P(x)$ within zone $i$ from~\zcref{eq:AFD_stencil}. This stencil is then evaluated at $x_{\pm 1/2}$ in order to get the cell interface values. It is also differentiated (analytically, by appropriately differentiating the Legendre polynomials in~\zcref{eq:AFD_k3_Lp}) and then evaluated at $x=0$ to get $\partial_x q$ at the centred point.

Again, all of the coefficients listed here can be constructed using computer algebra. The principle can then be extended to arbitrary order. 
The schemes constructed and recommended by~\cite{balsaraEfficientAlternativeFinite2024} use a combination of the third order scheme shown above, and the higher order scheme. This writes the stencil as a data-dependent interpolation between the high and low order stencils, such that for smooth data it limits to the high order stencil.

\subsection{Computing the correction term}
\label{sec:AFD_details_correction}

From the field reconstruction in~\zcref{sec:AFD_details_recon} we have enough information to construct the solution to the Riemann problem at the cell interfaces, and to construct the term needed for the correction, $\partial_q f \partial_x q$, at nodes. The final step is to compute the derivatives of this term necessary to compute the full correction in~\zcref{eq:AFD_flux3}, which is given by
\begin{equation}
    \label{eq:AFD_flux3_correction}
    \sum c_{2k} \left( \Delta x \right)^{2 k} \left[ \partial_x^{2 k-1} \left( \pdv{\mathbf{f}}{\mathbf{q}} \partial_x \mathbf{q}  \right)\right]_{i+1/2} \, .
\end{equation}

The terms that we want to compute are denoted $g(x)$ which are appropriate components of $\partial_q f \partial_x q$. The notation of~\cite{balsaraEfficientAlternativeFinite2024} uses the same Legendre polynomial basis as in~\zcref{sec:AFD_details_recon}, but the coordinate system is now centred on the cell interface at $x=+1/2$, rather than the node at $x=0$. The nodal values are then, for example,
\begin{equation}
    \label{eq:AFD_correction_location}
    g_0 = \left. \left( \pdv{f}{q} \pdv{q}{x} \right) \right|_{x=-1/2} \, .
\end{equation}

As in the reconstruction case in~\zcref{sec:AFD_details_recon}, the stencil $Q(x)$ is written as a convex combination of individual stencils
\begin{equation}
    \label{eq:AFD_correction_stencil}
    Q(x) = \sum \bar{\omega}_k Q_k(x) \, , \quad \sum \bar{\omega}_k = 1 \, .
\end{equation}
Because of the staggering of points with respect to the origin of the coordinate system, fewer individual stencils are required to compute the terms needed. For example, in the third order case, only the left-biased stencil $Q_1$ and the right-biased stencil $Q_2$ are required. The individual stencils are again found as interpolations of the nodal data. The change in the origin of the coordinate system means their explicit form changes, however. Again writing
\begin{equation}
    \label{eq:AFD_correction_stencils}
    Q_k(x) = \sum_{\ell=0} v^{(\ell)}_k L_\ell(x) \, ,
\end{equation}
the explicit coefficients for the third order scheme are
\begin{widetext}
\begin{equation}
    \label{eq:AFD_correction_k3_stencils}
    \begin{aligned}
        v^{(0)}_1 &= \tfrac{1}{12} \left( 8 g_0 - g_{-1} + 5 g_{1} \right) & v^{(1)}_1 &= \left( g_1 - g_{0} \right) & v^{(2)}_1 &= \tfrac{1}{2} \left( g_1 - 2 g_{0} + g_{-1} \right) \, , \\
        v^{(0)}_2 &= \tfrac{1}{12} \left( 8 g_1 - g_{2} + 5 g_{0} \right) & v^{(1)}_2 &= \left( g_1 - g_{0} \right) & v^{(2)}_2 &= \tfrac{1}{2} \left( g_2 - 2 g_{1} + g_{0} \right) \, .
    \end{aligned}
\end{equation}
\end{widetext}
The form of the individual smoothness indicators is identical to~\zcref{eq:AFD_k3_smoothness1}, as
\begin{equation}
    \label{eq:AFD_correction_k3_smoothness1}
    \beta_k = \left( v^{(1)}_k \right)^2 + \frac{13}{3} \left( v^{(2)}_k \right)^2 \, 
\end{equation}
whilst the global smoothness indicator becomes
\begin{equation}
    \label{eq:AFD_correction_k3_smoothness2}
    \tau =  \left| \beta_2 - \beta_1 \right| \, .
\end{equation}
The remaining steps, giving the construction of the nonlinear weights $\bar{\omega}_k$, are identical to~\zcref{sec:AFD_details_recon}.

This gives the piecewise polynomial reconstruction of the correction term as $Q(x)$. In the third-order case, it has been constructed as a second-order polynomial. 
We can therefore differentiate $Q(x)$ once and evaluate the result at $x=0$, obtaining the first correction term required in~\zcref{eq:AFD_flux3_correction}. Again, the systematic construction of higher-order interpolants and higher-order derivatives of the stencil can be completed using computer algebra.

\bibliographystyle{apsrev4-2}
\bibliography{refs}
\end{document}